\documentclass[11 pt,nofootinbib,notitlepage,eqsecnum]{revtex4-1}
\usepackage[english]{babel}
\usepackage{bbm}
\usepackage{mathtools} 
\usepackage{subfig}
\usepackage{graphicx,tikz}
\usetikzlibrary{arrows,shapes.geometric,positioning}
\usepgfmodule{shapes}
\usetikzlibrary{decorations.markings,decorations.text,decorations,patterns,calc}
\usepackage{pgfkeys}
\renewcommand{\bm}[1]{{\mathbf #1}}

\usepackage{amsmath,amssymb,amsthm,MnSymbol}

\usepackage{hyperref}

\newtheorem{lemma}{Lemma}

\newtheorem{theorem}{Theorem}
\newtheorem{definition}{Definition}

\newcommand{\startproof}{\noindent\textbf{Proof.} }
\newcommand{\finishproof}{\hfill $\blacksquare$ \\}

\newcommand{\dif}{\mathrm{d}}

\newcommand{\ee}[1]{\operatorname{e}^{#1}}
\newcommand{\mrix}[1]{\mathbbm{#1}}
\def\Re{\mathrm{Re}\,}

\def\tr{\mathrm{tr}}

\def\sgn{\mathrm{sgn}}

\newcommand{\ket}[1]{|#1\rangle}
\newcommand{\bra}[1]{\langle #1|}
\newcommand{\scal}[1]{\langle #1\rangle}

\newcommand{\hilbert}{\mathcal{H}}
\newcommand{\irrsl}{\hilbert_{(k,p)}}

\newcommand{\spinor}[2]{\begin{pmatrix}#1\\#2\end{pmatrix}}
\newcommand{\zweimatrix}[4]{\begin{pmatrix}#1&#2\\#3&#4\end{pmatrix}}

\def\R{\mathbb{R}}

\def\C{\mathbb{C}}
\def\CP{\mathbb{CP}}

\def\timeo{\mathcal{T}}

\newcommand{\slc}{\mathfrak{sl}(2,\C)}
\newcommand{\SL}{\mathrm{SL}(2,\C)}

\newcommand{\SO}{\mathrm{SO}}

\newcommand{\SU}{\mathrm{SU}}

\newcommand{\pauli}{{\boldsymbol\sigma}}
\newcommand{\rot}{{\bm L}}
\newcommand{\boost}{{\bm K}}
\def\rotoperator{\hat{\,\bm L}}
\def\boostoperator{\hat{\,\bm K}}

\newcommand{\dual}{\ast}
\newcommand{\geom}{\mathrm{geom}}

\newcommand{\geosimp}{\sigma}
\newcommand{\osimp}{\tilde{\sigma}} 
\def\orM{\mathring{\epsilon}}
\newcommand{\dummy}{\rule{0in}{0in}}

\def\SFreview{\cite{Baez:1999sr,Perez:2004hj,Perez:2012wv,Rovelli:2011eq}}
\def\LQGreview{\cite{ThiemannBook, Ashtekar:2004eh,RovelliBook,Sahlmann:2010zf}}
\def\LorentzEPRL{\cite{Pereira:2007nh, Engle:2008ev}}
\def\EuclideanEPRL{\cite{Engle:2007uq,Engle:2007qf,Engle:2007wy}}
\def\modernmodel{\cite{Engle:2007wy,Freidel:2007py,Kaminski:2009fm,Bahr:2010bs}}
\def\KKL{\cite{Kaminski:2009fm, Kaminski:2011bf, Kaminski:2009cc}}
\def\asymptoticL{\cite{Barrett:2009mw}}
\def\asymptoticE{\cite{Barrett:2009gg}}
\def\asymptotic{\cite{Barrett:2009mw,Barrett:2009gg}}
\def\jonProp{\cite{Engle:2011un}}
\def\jonCorr{\cite{Engle:2013psq}}  
\def\jonSec{\cite{Engle:2011ps}}
\def\jon{\cite{Engle:2011ps,Engle:2013psq,Engle:2012yg,Engle:2011un}}
\def\jonClass{\cite{Engle:2011ps, Engle:2013psq}}

\def\beq{\begin{equation}}
\def\eeq{\end{equation}}
\def\bq{\begin{equation*}}
\def\eq{\end{equation*}}

\newcommand{\propdec}{\mathrm{(+)}}

\newcommand{\Aprop}{A^{\propdec}_v}

\begin{document}
\title{The Lorentzian proper vertex amplitude: Classical analysis and quantum derivation}

\author{Jonathan Engle${}^{1,2}$}
\email{jonathan.engle@fau.edu}

\author{Antonia Zipfel${}^{2,3}$}
\email{antonia.zipfel@gravity.fau.de}

\affiliation{${}^1$ \mbox{Department of Physics, Florida Atlantic University, Boca Raton, Florida, USA}\\
${}^2$ \mbox{Institut f\"ur Quantengravitation, Department Physik Universit\"at Erlangen,}\\ Staudtstrasse 7, D-91058 Erlangen, Germany\\
${}^3$ \mbox{Instytut Fizyki Teoretycznej, Uniwersytet Warszawski, ul. Ho\.za 69, 00-681 Warsaw, Poland}}

\begin{abstract}
\centerline{\bf Abstract}

Spin foam models, an approach to defining the dynamics of loop quantum gravity, make use of the Plebanski formulation
of gravity, in which gravity is recovered from a topological field theory via certain constraints called simplicity constraints.
However,
the simplicity constraints in their usual form select more than just one gravitational sector as well as a degenerate sector.
This was shown, in previous work, to be the reason for the
``extra'' terms appearing in the semiclassical limit of the  Euclidean EPRL amplitude.
In this previous work, a way to eliminate the extra sectors, and hence terms, was developed, leading to the what was called the Euclidean proper vertex amplitude. 
In the present work, these results are extended to the Lorentzian signature, establishing what is called the Lorentzian proper vertex amplitude. This extension is non-trivial and involves a number of new elements since, for Lorentzian bivectors, 
the split into self-dual and anti-self-dual parts, on which the Euclidean derivation was based, is no longer available. 
In fact, the classical parts of the present derivation provide not only an extension to the Lorentzian case,
but also, with minor modifications, provide a new, more four dimensionally covariant derivation for
the Euclidean case.
The new elements in the quantum part of the derivation are due to the different structure of unitary representations of the Lorentz group.


\end{abstract}

\maketitle

\section{Introduction}

Spin foams {\SFreview} arise from a path integral approach to loop quantum gravity (LQG) {\LQGreview} and are based on the observation that general relativity in Plebanski's formulation \cite{Plebanski:1977zz} resembles $\SL$-BF-theory in four dimensions. BF-theories are named after the general form of their action,
$S_{BF}=\int_{\mathcal{M}} \mathcal{B}\wedge F$ where $\mathcal{B}$ is a Lie-algebra valued 2-form on a 4-manifold $\mathcal{M}$ and $F$ is the curvature of a connection $A$.
Since these theories are well understood and under good technical control, the general strategy in spin foams is first to discretize then quantize the BF action and afterwards implement the so-called \emph{simplicity constraints}
which reduce $BF$ theory to gravity. If we define the \textit{Plebanski 2-form} $B$ by
$2 \kappa \mathcal{B}=:B+\frac{1}{\gamma} \dual B$, with $\gamma$ the Barbero-Immirzi parameter
\cite{Barbero:1994ap, Immirzi1995}, $\kappa=8\pi G$, and
$\dual$ the hodge dual on the algebra, the solutions to this constraint fall into the five sectors:
\begin{enumerate}
\item[(I$\pm$)$\;$] $B = \pm e \wedge e$ for some $e$
\item[(II$\pm$)] $B = \pm \ast e \wedge e$ for some $e$
\item[(deg)] $B$ is degenerate
($\tr(\ast B \wedge B) = 0$)
\end{enumerate}
Of these, only the sectors (II$\pm$) correspond to gravity with co-tetrad $e$ and the usual Newton constant $G$.
In modern models {\modernmodel} the original constraints are replaced by a reformulation that is linear in the $B$-field, 
and which excludes solutions of type (I$\pm$). Nevertheless, the question which of the remaining \emph{Plebanski sectors} 
are still present  and how they affect the resulting amplitudes is seldom addressed in the literature. For the Euclidean EPRL-model {\EuclideanEPRL} this issue was examined for the first time in {\jonClass} where it was found that the sectors (deg) and (II$\pm$) are still present in their entirety. It was further argued that a certain mixing of these sectors is responsible for the appearance of multiple undesired terms in the asymptotic limit of the model. More specifically, in their important work {\asymptoticE} Barrett, Dowdall, Fairbrain, Gomes  and Hellmann discovered that in the large spin-limit the EPRL-amplitude is governed by multiple terms,
each consisting in an exponential of the Regge action, that is, the discrete action of gravity, rescaled by different coefficients
and different signs, rather then by the one exponential term one would expect.
This leads to unphysical equations of motion dominating in this limit when multiple 4-simplices are considered
\cite{Engle:2012yg, clrrr2012} and may even be the source of bubble divergences \cite{clrrr2012}.
Also, the authors of \cite{Zipfel3} argued that the appearance of these multiple sectors might be responsible for the difficulties arising when trying to construct a Rigging-map for the canonical theory from the KKL-model {\KKL}, a generalized version of the EPRL-model.
All these problems might therefore be solved by imposing an additional quantum constraint involving orientation and Plebanski sectors of the $B$-fields and eliminating the undesired sectors and hence terms. 
Such a quantum constraint has already been used to modify the Euclidean model {\jon} leading to what was called the
\emph{proper vertex amplitude}. 

Of course, the Euclidean model is just a testing ground for methods to be used in the Lorentzian theory.
The aim of this paper is to extend the analysis of {\jonClass} 
to the Lorentzian case, and to derive a Lorentzian proper vertex amplitude. 
The outcomes are similar but not identical to those in the Euclidean case: 
In the Lorentzian EPRL model, all three sectors (II$\pm$) and (deg) are present and generate the various terms in the asymptotics. The undesired sectors and hence terms can be excluded by imposing an analoguous quantum constraint 
yielding a Lorentzian proper vertex amplitude that is both $\SU(2)$- and $\SL$-covariant. The most obvious differences from the Euclidean case arise from the fact that the $\SL$ algebra does not decompose into self-dual and anti-self-dual parts, and has very different unitary representations, forcing 
the presentation of the classical analysis, and the details of the quantum analysis, to be quite different from that used in 
the Euclidean paper.  
In particular, in the classical analysis, one is forced to use structures which make the four dimensional geometry more manifest throughout presentation.
The resulting structure of the classical analysis, in addition to providing an extension to the Lorentzian case, 
provides, with minor modification, a new derivation for the Euclidean signature,
so that this part of the presentation is more general than that in {\jon}.

The paper is organized as follows. 
Section \ref{sec:class} presents the classical analysis part of the paper.
Specifically, in section \ref{ssec:disc}
the Lorentzian bivector geometries will be reviewed based on {\asymptoticL}.
In section \ref{sec:pleb} these discrete geometries will be related to continuous 2-forms, and a characterization of  
the different Plebanski sectors in terms of the discrete data is derived.
Finally, in section \ref{sec:Einstein-Hilbert}, we analyze the classical constraint restricting to the Einstein-Hilbert sector closely following the ideas of {\jon}.
In the third section we will recall the construction of the Lorentzian EPRL-model {\LorentzEPRL} and its asymptotic limit, 
and interpret each term in this limit as arising from a corresponding Plebanski sector. Finally, in section \ref{sec:proper} we will propose a new proper vertex amplitude and analyze its symmetry properties. The semiclassical properties of this new amplitude will be studied in the companion paper \cite{evz2015}, and its graviton propagator is studied in the upcoming paper
\cite{cev2015}.

\section{Classical analysis}
\label{sec:class}

The spin foam approach is based on a discretized classical theory and for this reason
the path integral should be peaked on discrete geometries in the semiclassical limit.
Even though for a complete semiclassical picture one should take arbitrarily fine triangulations into account involving an arbitrary number of simplices (see e.g. \cite{Hellmann:2012kz}), we will only consider the amplitude associated to a single 4-simplex here,
as this is simplest and is sufficient to identify the different Plebanski sectors in the asymptotic limit.

The spin foam path integral defines transition amplitudes for states of quantum geometry on its boundary.
In the case of a single 4-simplex $\geosimp$ this means that the associated amplitude describes quantum evolution
from one part of its boundary $\partial\geosimp$ to the other, each triangulated by tetrahedra. Hence, we will assume that all tetrahedra are space-like\footnote{For an extension of the EPRL-model to incorporate also time-like tetrahedra see \cite{Conrady:2010kc,Conrady:2010vx}.}.

\subsection{Discrete Lorentzian geometry}
\label{ssec:disc}

\subsubsection{Discrete simplicial geometry}
\label{ssec:discrete}
For the purpose of discussing the data, geometry, and fields associated to a single 4-simplex,
we use an oriented manifold $\mathcal{M}$ 
isomorphic to $\R^{3,1}$ equipped with the Minkowski metric $\eta=\mathop{diag}(-,+,+,+)$.
\begin{definition}[Geometric 4-simplex]
A geometric 4-simplex $\geosimp$ is the convex hull of 5 points in $\mathcal{M}$, called vertices, not all of which lie in the same 3d affine hyperplane.
\end{definition}
\begin{definition}[Numbered 4-simplex]
A numbered 4-simplex is a geometric simplex together with an (arbitrary) assignment of labels $0,\cdots,4$ to its five vertices. Each tetrahedron $\tau_a$ is then labeled by the vertex $a$ it does not contain and the common triangle
$\Delta_{ab}$ of tetrahedron $\tau_a$ and $\tau_b$ is labeled by the pair $(ab)$.
\end{definition}
Note, the numbering of the vertices induces an orientation, which we call `discrete', on the simplex $\geosimp$ independent of the orientation of $\mathcal{M}$: By fixing the vertex $0$ the ordered quadruplet $(\vec{01},\vec{02},\vec{03},\vec{04})$ of vectors at $0$ tangent to the edges joining $0$ and $a=1,2,3,4$ defines an orientation.
Of course there are also other ways to associate an orientation to each numbering.
In the subsequent discussion the concrete convention how to relate numberings and orientation is not of interest;
however, once it is established, the so-defined discrete orientation either agrees or disagrees with the orientation of the manifold.
At certain points we will restrict to numberings which agree with the orientation of the manifold;
when we do this we thereby endow the numbering with information about orientation which is not given elsewhere in the
discrete data.
\begin{definition}[Ordered 4-simplex]
\label{def:ordered}
An ordered 4-simplex is a numbered 4-simplex in an oriented manifold $\mathcal{M}$ whose discrete orientation coincides with the orientation of $\mathcal{M}$.
\end{definition}

\subsubsection{Bivector geometry}
\label{sec:bivector}

Simple bivectors, elements in $\Lambda^2(\R^{3,1})$ of the form $\Sigma=v\wedge w$,
are naturally associated to flat triangles because they are fixed uniquely by two edge vectors $v,w\in\R^{3,1}$.
If the plane spanned by $v,w$ is space-like, time-like, or null, $\Sigma$ is respectively called space-like, time-like, or null.
The norm of $\Sigma$, 
\begin{align*}
\Sigma^2:= \frac{1}{2} \Sigma^{IJ}\Sigma_{IJ}
\end{align*}
in these three cases is respectively positive, negative, or zero.
Here and throughout this paper, Minkowski space indices
$I,J,K \dots$ are raised and lowered with the Minkowski metric $\eta$.

Alternatively, the information in a triangle can be encoded in the bivector $\ast v\wedge w$. It is not hard to see that this gives again a simple bivector, $N\wedge N' \propto\ast v\wedge w$, where now $N$ and $N'$
are appropriately choosen \textit{normals} to the triangle. 
Specifically, to a triangle $\Delta_{ab}$ in a 4-simplex $\geosimp$ with area $A(\Delta_{ab})$, we associate 
the time-like bivector
%
%
\begin{align}
\label{eqn:geomB}
B^{\geom}_{ab}(\geosimp)=-A(\Delta_{ab})\frac{N_a\wedge N_b}{|N_a\wedge N_b|}
\end{align}
where $N_a$ is the outward pointing normal to the $a^{\text{th}}$ tetrahedron, $|B|:= |B^2|^{1/2}$,
and the sign is chosen for later convenience\footnote{Specifically
so that the sign in (\ref{eqn:contB}) is positive.
%
%
}.
In the following such bivectors are called \emph{geometric}\footnote{In
{\asymptotic} a geometric bivector $\tilde{B}^{\geom}_{ab}$ is defined as the bivector spanning the plane of $\Delta_{ab}$, i.e.
$\ast\tilde{B}^{\geom}_{ab}= B^{\geom}_{ab}$. Here, we have chosen instead to be consistent with the convention used in \cite{bhnr2004} and {\LorentzEPRL}.}. As a 4-simplex has 10 triangles this gives a set of 20 bivectors such that $B_{ab}^{\geom}=- B_{ba}^{\geom}$.
\begin{definition}[Discrete Plebanski field]
A set of time-like simple bivectors $\{B_{ab}\},\, a,b=1,\dots,5$  is called a discrete Plebanski field if
\begin{enumerate}
\item $B^{IJ}_{ab}=-B^{IJ}_{ba}$ (\textit{orientation})
\item $\sum\limits_{b:b\neq a} B^{IJ}_{ab}=0$ (\textit{closure})
\end{enumerate}
\end{definition}
\begin{definition}[Bivector geometry]
A discrete Plebanski field is a weak bivector geometry if
\begin{enumerate}
\item $\forall a\;\;\exists\, N^I_a\in\R^{3,1}$ such that $N_{a I}(\ast B_{ab})^{IJ}=0,\, \forall b\neq a$ (\textit{linear simplicity})
\item $\tr(B_{ab}[B_{ac},B_{ad}])\neq0$ (\textit{tetrahedron non-degeneracy})
\end{enumerate}
If additionally
$\{B_{bc}\}_{b,c \neq a}$
spans $\Lambda^2(\R^{3,1})$ for any $a$
(full non-degeneracy) then it is called a bivector geometry.
%
%
\end{definition}
As Barrett and Crane showed in their seminal work \cite{Barrett:1997gw}, any 4-simplex in $\R^4$ determines an Euclidean bivector geometry in $ \Lambda^2(\R^4)$ and any Euclidean bivector geometry defines a 4-simplex unique up to translation and inversion.
At no point in their proof is the Euclidean metric used, so that the same proof applies to the case of a Lorentzian
signature metric.
Furthermore, the proof does not make use of any predefined orientation so that the resulting simplices are numbered but not necessarily ordered. We thus have:
 \begin{theorem}[Bivector geometry theorem]
\label{th:bivector_geometry}
 The bivectors  $\{B_{ab}^{\geom}\}\subset\Lambda^2(\R^{3,1})$ associated to a numbered 4-simplex with space-like boundary form a bivector geometry. Vice versa, any bivector geometry determines a 4-simplex $\geosimp$ of the above type, unique up to translation and inversion, such that  $B_{ab}=\mu\, B^{\geom}_{ab}(\geosimp)$ for some unique $\mu = \pm 1$.
 \end{theorem}
One can separate the information in a bivector geometry into \textit{boundary data} describing
each tetrahedron separately, and \textit{parallel transports} describing how the tetrahedra fit together.
Keeping in mind the later use in quantum theory, it will be convenient to describe the parallel transports using $\SL$
elements.
Each $\SL$ element is associated to an element of the proper orthochronous Lorentz group $\SO^+(3,1)$,
using the isomorphism
\begin{align}
\label{eqn:rho}
\rho:\R^{3,1}\to \mathbb{H} \qquad v\mapsto v^0\,\mathrm{Id}+\mathbf{v}^i\pauli_i~,
\end{align}
where $\mathbb{H}$ is the space of 2x2 Hermitian matrices and $\pauli_i$ are the Pauli matrices.
A given $\Lambda \in \SL$ then corresponds to the unique  $\widehat{\Lambda}\in\SO^+(3,1)$ for which
\begin{align}
\label{eqn:2to1}
\rho(\widehat{\Lambda} v)=\Lambda \rho(v)\Lambda^\dagger.
\end{align}
%
%
Given any weak bivector geometry $\{B_{ab}^{IJ}\}$, one then has that there exist areas $\{A_{ab}=A_{ba}\}$,
3-vectors $\{\bm{n}_{ab}\}$, and $\SL$ elements $\{X_a\}$
such that\footnote{This
equation fixes the meaning of $X_a$, $A_{ab}$, and $\bm{n}_{ab}$.
In particular the minus sign is a choice of convention in the meaning of $\bm{n}_{ab}$,
chosen so that the coefficient relating $\bm{n}_{ab}$ and $\bm{L}_{ab}$ in equation (\ref{nLrelation})
is positive.
%
%
}
%
%
\begin{align}
\label{Bdef}
B_{ab} = - A_{ab} \widehat{X}_a \triangleright \timeo \wedge (0, \bm{n}_{ab})
\end{align}
where $\mathcal{T}:= (1,0,0,0)$ and $\triangleright$ denotes the action of $\SO(3,1)$
on $\Lambda^2(\R^{3,1})$, that is, $(\widehat{X} \triangleright b)^{IJ}:= \widehat{X}^I{}_K \widehat{X}^J{}_L b^{KL}$.
The `boundary data' $\{A_{ab}, \bm{n}_{ab}\}$ determines the geometry of each
tetrahedron individually, with the group elements $\widehat{X}_a$ telling how to rotate these
tetrahedra so they `glue' together to form a 4-simplex.
%
%
Specifically, the \textit{closure} condition on $\{B_{ab}\}$ becomes
\begin{align}
\label{eqn:bdclosure}
\sum_{b:b \neq a} A_{ab} \bm{n}_{ab} = 0,
\end{align}
and the tetrahedron non-degeneracy condition becomes
\begin{align}
\label{btnondeg}
\{\bm{n}_{ab}\}_{b:b\neq a}\text{ span }\R^3,
\end{align}
which are the necessary and sufficient conditions for there to exist a tetrahedron $t_a$ in $\R^3$
with areas $\{A_{ab}\}_{b:b\neq a}$ and outward pointing normals $\{\bm{n}_{ab}\}_{b:b\neq a}$
to each face \cite{Minkowski:poly}.
If we identify $\R^3$ with the hypersurface in $\R^{3,1}$ orthogonal to $\mathcal{T}$,
the simple time-like bivectors
associated to each face of $t_a$ are
\begin{align}
\label{eqn:b_ab}
b_{ab} := - A_{ab} \mathcal{T} \wedge (0, \bm{n}_{ab}).
\end{align}
The bivectors $B_{ab}$ are then obtained by rotating these by $\widehat{X}_a$.
Let $\tau_a$ denote the rotated tetrahedron $\widehat{X}_a t_a$.
Each tetrahedron $t_a$
lies in $\R^3$ and hence has $\pm \mathcal{T}$ as its normal.
Consequently, the rotated tetrahedron $\tau_a$ associated
to the bivectors $\{B_{ab}\}_{b:b\neq a}$ -- that is, the $a$th tetrahedron of the 4-simplex $\geosimp$ reconstructed in theorem
\ref{th:bivector_geometry} -- has
\begin{align}
\label{Neq}
N_a = \pm F_a:= \pm \widehat{X} \mathcal{T}
\end{align}
as its normal, where we let the sign $\pm$ be chosen separately for each $a$ such that $N_a$
is outward pointing.
This is consistent with the fact
that the simple time-like bivectors $b_{ab}$ satisfy linear simplicity using $\pm \mathcal{T}$ as normal,
\begin{align}
\label{gflinsimp}
\timeo_I (\dual b_{ab})^{IJ} = 0,
\end{align}
while the bivectors $\{B_{ab}\}_{b:b\neq a}$ satisfy linear simplicity
using $N_a = \pm \widehat{X}_a \mathcal{T}$:
\begin{align}
\label{linsimpsat}
N_{aI} \ast B_{ab}^{IJ}
= (\widehat{X}_a\timeo)_I \ast B_{ab}^{IJ}
= 0.
\end{align}
In terms of $\{A_{ab}, \bm{n}_{ab}, X_a\}$, the orientation constraint on $\{B_{ab}\}$ becomes
\begin{align}
\label{orient}
\widehat{X}_a \triangleright b_{ab} = - \widehat{X}_b \triangleright b_{ba}
\end{align}
Finally, the full non-degeneracy condition is equivalent to a condition only on the group elements $\{X_a\}$.
In preparation for stating this condition, note that,
given a solution $\{X_a\}$ to (\ref{orient}), for any set of signs $\epsilon_a$ and any $Y \in \SL$,
$\{X'_a := \epsilon_a Y X_a\}$ is also a solution.  We call any such pair of solutions `equivalent'
because
(1.) the signs
$\epsilon_a$ do not affect $\widehat{X}_a$ and hence do not affect the final reconstructed $B_{ab}$,
and (2.) the rotation $Y$ just effects an overall global $SO(3,1)$ rotation,
and we write $\{X_a\}\sim\{X_a'\}$.
\begin{lemma}
\label{lem:non-deg}
The bivectors $\{B_{ab}=\widehat{X}_a\triangleright b_{ab}\}$, constructed from a given set of closed, non-degenerate boundary data
and $X_a$ satisfying orientation, are degenerate iff $\{X_{a}\}\sim\{U_a\}\subset\SU(2)$.
\end{lemma}
{\startproof
Assume that $\{X_{a}\}$ is equivalent to a subset in $\SU(2)$.
Then $\widehat{X}_a = \widehat{Y} \widehat{X}_a'$ for some
$\widehat{Y} \in \SO^+(3,1)$ and $\{\widehat{X}_a'\} \subset \SU(2)$, so that
$\widehat{X}_a\timeo=\widehat{Y}\timeo$. But by (\ref{linsimpsat}) this means that
\bq
0=(\widehat{X}_a\timeo)_I(\ast B_{ab})^{IJ}=(Y\timeo)_I(\ast B_{ab})^{IJ}
\eq
for all $a$ and $b$, whence $\{B_{ab}\}$ is degenerate.

Suppose conversely that $\{B_{ab}\}$ is degenerate.  By lemma 3 of {\asymptoticE}
(which applies unaltered to the Lorentzian case {\asymptoticL}) it follows that
all future pointing normals $F_a=\widehat{X}_a\timeo$ are proportional and hence equal to some 4-vector $F$.
Let $Y$ be any $\SL$ element such that $\widehat{Y} \timeo = F$, and let
$X_a':= Y^{-1} X_a$, so that $X_a = Y X_a'$.
Then $\widehat{X}_a' \timeo = Y^{-1} F = \timeo$,
which from (\ref{eqn:2to1}) implies
$\{X_a'\} \subset \SU(2)$, so that $X_a$ is equivalent to a subgroup of $\SU(2)$.
\finishproof}
\noindent Due to this lemma, if a set of group elements $\{X_a\}$ is equivalent to a subset of $\SU(2)$, we shall call it
\textit{degenerate}.

To summarize, any weak bivector geometry $\{B_{ab}\}$ can be written in the form (\ref{Bdef})
for some $\{A_{ab}, \bm{n}_{ab}, X_a\}$.
Conversely, given any set of data  $\{A_{ab}, \bm{n}_{ab}, X_a\}$, the reconstructed bivectors
 (\ref{Bdef}) will form a weak bivector geometry iff (\ref{eqn:bdclosure}), (\ref{orient}), and (\ref{btnondeg})
are satisfied, and will form a  bivector geometry iff $\{X_a\}$ is non-degenerate.

Given boundary data $\{A_{ab}, \bm{n}_{ab}\}$, one can also ask whether there \textit{exist} group elements
$\{X_a\}$ so that (\ref{Bdef}) form a weak bivector geometry, how many \textit{inequivalent}
such choices of group elements $\{X_a\}$ exist and whether they are degenerate.
This question has been analyzed in {\asymptoticL} and leads to the following classification.
Suppose $\{A_{ab}, \bm{n}_{ab}\}$ satisfy closure and tetrahedron non-degeneracy.
\begin{enumerate}
\label{classi}
\item
If the boundary data correspond to the boundary of a non-degenerate Lorentzian 4-simplex, there will exist two inequivalent solutions
$X_a$ to the orientation constraint, neither of which is degenerate.
\item
If the boundary data correspond to the boundary of a non-degenerate
Euclidean 4-simplex, there will exist two inequivalent solutions $X_a$
to the orientation constraint, \textit{both of which} are degenerate\footnote{In this case, the two solutions are equivalent to two
solutions $\{U_a^+\}$ and $\{U_a^-\}$ in $SO_\mathcal{T}(3)$.
These can be used to reconstruct the four dimensional bivectors of the relevant Euclidean 4-simplex via
$B_{ab}^{\text{Eucl}}:= [(U^+_a, U^-_a)] \triangleright b_{ab}$
where $\triangleright$ denotes the adjoint action of
$SO(3) \times SO(3)/\mathbb{Z}_2 \cong SO(4)$ on $\mathfrak{so}(4)$.}.

\item
If the boundary data neither correspond to the boundary of a non-degenerate Lorentzian 4-simplex, nor to the boundary of a
non-degenerate Euclidean 4-simplex, then it forms what is called a \textit{vector geometry}
{\asymptotic}, and, up to equivalence, there exists only one solution $X_a$, which is degenerate.

\end{enumerate}
%
%
Note that both of the last two cases correspond to \textit{degenerate}
weak bivector geometries from the Lorentzian point of view.

The first two cases furthermore share a notable property which will be important later on when defining coherent
states corresponding to given boundary data.
Namely, in the first two cases, not only are the bivectors $B_{ab}$ and $B_{ba}$ equal and opposite,
but the corresponding triangles also \textit{have the same shape}
so that the reconstructed tetrahedra glue together
to provide a consistent Regge geometry of the boundary of the 4-simplex.  This is a property which can
be coded in the boundary data alone, and such boundary data is called \textit{Regge-like}.
\begin{definition}[Regge-like boundary data]
A set of closed non-degenerate boundary data $\{\bm{n}_{ab}, A_{ab}\}$ is called Regge-like if the tetrahedra $t_a$ defined by $\{\bm{n}_{ab},A_{ab}\}_{b:b\neq a}$ glue together to form a consistent Regge geometry on the boundary of a 4-simplex.
\end{definition}
For such boundary data there exists a unique set of $\SO(3)$ elements $\widehat{g}_{ab}$, called \textit{gluing maps},
such that
\begin{align}
\label{eqn:gluing}
\widehat{g}_{ab} \bm{n}_{ab} = - \bm{n}_{ba}
\text{ and } \widehat{g}_{ab} t_a = t_b .
 \end{align}

So far we have not discussed degenerate boundary data ---
that is, for which tetrahedron non-degeneracy fails.
For this kind of data the geometric picture is fundamentally different since the normals $\bm{n}_{ab}$ can no longer be interpreted as normals of the triangles but are edges of a quadrilateral. Such configurations are analyzed in greater detail in \cite{Barrett:2002ur} but will be excluded here since the methods of {\asymptoticL} can not be applied to determine the phase in the asymptotic formula.

\subsection{Plebanski Sectors}
\label{sec:pleb}

\subsubsection{From the continuum to the discrete}

In the Plebanski-Holst formulation of gravity \cite{Plebanski:1977zz, Holstaction, Capovilla:2001zi,
Engle:2007qf, Freidel:2007py, Engle:2007wy} the usual Einstein-Hilbert action is replaced by a BF-type action with an additional constraint:
 \begin{align}
\label{BFaction}
S =- \frac{1}{2\kappa}\int_{\mathcal{M}} \tr[(B+ \frac{1}{\gamma} \ast B) \wedge F[A]+\lambda_{IJKL}B^{IJ}\wedge B^{KL}]
\end{align}
where $B$ is a $\slc$-valued 2-form on an oriented space-time $\mathcal{M}$ and $F[A]$ is the curvature of a $\slc$-valued connection $A$. Furthermore, $\kappa$ denotes the gravitational constant $8\pi G$, $\gamma\in\R$ the Barbero-Immirzi parameter and $\lambda_{IJKL}$ a tensor satisfying $\lambda_{IJKL}\epsilon^{IJKL}=0$. Extremization of the above action with respect to the Lagrange multiplier $\lambda$ leads to the simplicity constraint
\begin{align}
\label{eqn:simpl}
B^{IJ}\wedge B^{KL}=\frac{1}{4!}\epsilon^{IJKL}\epsilon_{MNPQ}\,B^{MN}\wedge B^{PQ}
\end{align}
whose solutions fall into the five \textit{Plebanski sectors}:
\begin{enumerate}
\item[(I$\pm$)] $B = \pm e \wedge e$ for some $e^I_\mu$
\item[(II$\pm$)] $B = \pm \ast e \wedge e$ for some $e^I_\mu$
\item[(deg)] $B$ is degenerate
($\tr(\ast B \wedge B) = 0$)
\end{enumerate}
However, only if $B$ is in sector (II$\pm$) do the equations of motions of \eqref{BFaction} describe gravity
with the usual Newton constant and the action reduce to
the Holst action \cite{Holstaction} of Loop Quantum Gravity up to a sign.
Before quantizing the model the action is discretized and the continuous fields are replaced by
\begin{align}
\label{eqn:discB}
B_{\Delta}:=\int_{\Delta}B
\end{align}
where $\{\Delta\}$ are arbitrarily oriented triangles of a simplicial triangulation of $\mathcal{M}$.
In order to demonstrate how this discrete data are related to the discrete Plebanski fields of the last subsection, we focus on
a single 4-simplex $\geosimp$ in the triangulation. Number the vertices of $\geosimp$ from $0$ to $4$ such that
$\geosimp$ becomes an \textit{ordered 4-simplex}, i.e. in such a way that the numbering codes the orientation.
As in the last subsection, let $\tau_a$ denote the tetrahedron labeled by the number $a$ of the vertex opposite it, oriented as part of the boundary of $\geosimp$,  and let $\Delta_{ab}$ denote the triangle shared by $\tau_a$ and $\tau_b$, with orientation as part of the boundary of $\tau_a$.  We then define
\begin{align}
\label{simpdisc}
B_{ab}:= \int_{\Delta_{ab}} B.
\end{align}
These are the data which will be identified with a (weak) bivector geometry.
However, this identification is exactly possible only once the infinite number of degrees of freedom in the choice of continuum 2-form is restricted in some way. The restriction we choose is that the continuum 2-form should
be constant within $\geosimp$. Specifically, we fix a flat connection $\partial$ on $\geosimp$ which is
adapted to $\geosimp$ in the sense that, relative to the affine structure defined by $\partial$,
$\geosimp$ is the convex hull of its vertices.
We then restrict to continuum form fields $B$ which are constant with respect to $\partial$ in $\geosimp$.

The connection $\partial$ not only induces an isomorphism among all the tangent spaces of $\geosimp$, but also provides a natural
isomorphism between each tangent space and the space of constant vector fields on $\geosimp$.  Using this, one can give
a more explicit expression for the orientation of $\Delta_{ab}$ via the following definition.
\begin{definition}[Oriented triangle]
\label{orientedTriangle}
Let $\epsilon^{\alpha\beta\gamma\delta}$ be any inverse volume form\footnote{In
this subsection, greek indices refer to space-time while I,J,K, \dots are internal (algebra) indices.
Furthermore, it is here most convenient to define the orientation of a manifold by an equivalence class of \textit{inverse}
volume forms \jonProp.}
in the equivalence class defining the orientation of
$\mathcal{M}$ and let $N_a$ and $N_b$ be any outward pointing covariant normals to the tetrahedra $\tau_a$ and $\tau_b$ containing $\Delta_{ab}$. The inverse 2-form
\begin{align}
\label{eqn:indOr}
\epsilon^{\alpha\beta}_{[ab]}:=\epsilon^{\alpha\beta\gamma\delta}(N_a)_{\gamma} (N_b)_{\delta}
\end{align}
is non-zero, tangent to $\Delta_{ab}$ and unique up to rescaling by a positive function,
thereby fixing the orientation of the triangle.
\end{definition}
Note that, in this definition, the 4-simplex need not be ordered, so that $\epsilon^{\alpha\beta}_{[ab]}$
can be defined for any numbered simplex.
Furthermore, the orientation of the triangle is anti-symmetric under the exchange of the normals i.e. $\epsilon^{\alpha\beta}_{[ab]}=-\epsilon^{\alpha\beta}_{[ba]}$. For this reason, the data \eqref{simpdisc}
obey orientation. 

\subsubsection{From the discrete to the continuum}

To interpret a given weak bivector geometry in terms of Plebanski sectors, the continuum 2-form has to be reconstructed.
Lemma 1 in {\jonSec} tells us that for any discrete Plebanski field $\{B_{ab}\}$
and \emph{any choice} of a numbered 4-simplex $\geosimp$ in $\mathcal{M}$ there exist a \emph{unique} constant 2-form $B_{\mu\nu}(\{B_{ab}\},\geosimp)$ such that
\begin{align}
\label{disccont}
B_{ab}^{IJ} = \int_{\Delta_{ab}(\geosimp)} B^{IJ}(\{B_{a'b'}\},\geosimp)
\end{align}
where the orientation of triangle $\Delta_{ab}(\geosimp)$ is defined as in definition \ref{orientedTriangle}.
In general, the resulting continuum 2-form depends on the numbered 4-simplex chosen, though, as we shall see,
certain properties are independent of this choice. It will be important to consider how the resulting continuum 2-form
changes if we replace $\geosimp$ by its parity reversal $P\geosimp$.
The orientation $\epsilon^{\alpha\beta}_{[ab]}(P\geosimp)$ of $\Delta_{ab}(P\geosimp)$ is related to that of
$\Delta_{ab}(\geosimp)$ by
\begin{align}
\label{epsparity}
\epsilon^{\alpha\beta}_{[ab]}(P\geosimp)=\epsilon^{\alpha\beta\gamma\delta} (P^*N_a)_{\gamma} (P^*N_b)_{\delta}
=-P_{\ast}\epsilon^{\alpha\beta}_{[ab]}(\geosimp)
\end{align}
where we used that the pushforward $P_{\ast}$ on $\epsilon^{\alpha\beta\gamma\delta}$ yields
$P_{\ast}\epsilon =-\epsilon$. Applying this to the form integral yields
\begin{align*}
\int_{\Delta_{ab}(P\geosimp)}[-P^*B(\{B_{a'b'}\},\geosimp)]
= \int_{P\Delta_{ab}(\geosimp)} P^*B(\{B_{a'b'}\},\geosimp)
= \int_{\Delta_{ab}(\geosimp)} B(\{B_{a'b'}\}, \geosimp) = B_{ab}
\end{align*}
where (\ref{epsparity}), the diffeomorphism covariance of the form integral, and equation (\ref{disccont})
respectively have been used.
Comparing the final result with equation (\ref{disccont}), proves\footnote{For an alternative proof see Lemma 3 in {\jonCorr}}:
\begin{lemma}
\label{lem:parity}
Given any discrete Plebanski field $\{B_{ab}\}$ and any numbered 4-simplex $\geosimp$,
\begin{align}
\label{eqn:parity_cont}
B_{\mu \nu}(\{B_{ab}\},P\geosimp)=-P^{\ast}B_{\mu \nu}(\{B_{ab}\},\geosimp)~.
\end{align}
\end{lemma}
The continuum 2-form determined by equation \eqref{disccont}
defines a dynamical orientation which may or may not agree with $\epsilon$ and can be measured by
\begin{align}
\label{omegadef}
\omega(B_{\mu\nu}) :=\mathrm{sgn}\left(\epsilon^{\alpha\beta\gamma\delta}\,\epsilon_{IJKL} \,
B_{\alpha\beta}^{IJ}\, B_{\gamma\delta}^{KL}\right)~.
\end{align}
Similarly we can define a function $\nu(B_{\mu\nu})$ measuring the Plebanski sector, such that $\nu(B_{\mu\nu})$ equals $\pm1$ if $B_{\mu\nu}$ is in Plebanski sector (II$\pm$) and vanishes if $B_{\mu\nu}$ is degenerate.

Note that both $\omega$ and $\nu$
are invariant under the action of all orientation preserving diffeomorphisms but change sign when replacing $\geosimp$
by $P\geosimp$. This follows from Lemma \ref{lem:parity},
together with the fact that multiplication by a minus flips the Plebanski sectors while $P$ changes the orientation. That is,
\begin{align}
\label{eqn:omegaP}
\omega\left(B_{\mu\nu}(\{B_{ab}\},P\geosimp)\right)=- \omega\left(B_{\mu\nu}(\{B_{ab}\},\geosimp)\right)
\;\;\text{ and }\;\;
\nu\left(B_{\mu\nu}(\{B_{ab}\},P\geosimp)\right)=- \nu\left(B_{\mu\nu}(\{B_{ab}\},\geosimp)\right).
\end{align}
It follows that the Plebanski sector and orientation are ill-defined as functions of the discrete variables unless a restriction
is placed on the class of 4-simplices.
\footnote{However, the product of $\omega$ and $\nu$ is well-defined, which already hints that orientation and Plebanski sector can not be treated separately.} 
The most natural convention is to define $\omega$ and $\nu$ with respect to an \textit{ordered} 4-simplex, that is,
\begin{align}
\label{algfns}
 \nu(\{B_{ab}\}):= \nu\left(B_{\mu\nu}(\{B_{ab}\},\osimp)\right)\quad\text{and}\quad
 \omega(\{B_{ab}\}):= \omega\left(B_{\mu\nu}(\{B_{ab}\},\osimp)\right)
\end{align}
for any \textit{ordered} 4-simplex $\osimp$.
This convention endows the ordering of the tetrahedra with the information about the background orientation necessary to extract $\omega$ and $\nu$.

Once one makes this restriction to ordered 4-simplices $\osimp$,
as in the Euclidean case (see lemma 2 in {\jonSec}),
$\omega(B_{\mu\nu}(\{B_{ab}\}, \osimp))$ and $\nu(B_{\mu\nu}(\{B_{ab}\}, \osimp))$
are in fact independent of the choice of ordered 4-simplex $\osimp$ used.
This can be seen from the fact that the only background structures on $\mathcal{M}$ used in the
construction of the 2-form $B_{\mu\nu}(\{B_{ab}\},\osimp)$ are $\partial$, which is equivalent to an affine structure, and an orientation. The symmetry group of these background structures is the group of proper, i.e. orientation preserving, affine transformations. It follows that  $B_{\mu\nu}(\{B_{ab}\},\osimp)$ is covariant under this group, so that for any proper affine
transformation $\phi:\mathcal{M} \rightarrow \mathcal{M}$
\begin{equation*}
B_{\mu\nu}(\{B_{ab}\}, \phi \cdot \osimp) = \phi\cdot B_{\mu\nu}(\{B_{ab}\}, \osimp) :=  (\phi^{-1})^* B_{\mu\nu}(\{B_{ab}\}, \osimp)
\end{equation*}
where $\phi \cdot B_{\mu\nu}:= (\phi^{-1})^* B_{\mu\nu}$ is the natural left action of $\phi$ on a 2-form.
But given any two ordered 4-simplices $\osimp$ and $\osimp'$ there exist a proper affine transformation $\phi$ mapping the vertices of $\osimp$ to those of $\osimp'$ and therefore
\bq
B_{\mu\nu}(\{B_{ab}\},\osimp')=B_{\mu\nu}(\{B_{ab}\},\phi \cdot \osimp)=(\phi^{-1})^* B_{\mu\nu}(\{B_{ab}\},\osimp)~.
\eq
Since $\phi$ defines a particular orientation preserving diffeomorphism, it follows that
$\omega (B_{\mu\nu}(\{B_{ab}\}, \osimp')) = \omega(B_{\mu\nu}(\{B_{ab}\}, \osimp))$
and
$\nu(B_{\mu\nu}(\{B_{ab}\}, \osimp')) = \nu(B_{\mu\nu}(\{B_{ab}\}, \osimp))$, so that $\omega$ and $\nu$
are independent of the ordered 4-simplex used, proving the claim.

One can go further: The resulting functions $\omega(\{B_{ab}\})$ and $\nu(\{B_{ab}\})$
are also independent of the two background structures chosen on $\mathcal{M}$, namely the choice of flat connection
$\partial$ and the choice of a fixed orientation $\epsilon$.
To prove independence of $\partial$, let us begin by including $\partial$ explicitly as an argument,
so that we write $\omega(\{B_{ab}\}, \partial)$, $\nu(\{B_{ab}\}, \partial)$.
For the intermediate steps of the proof,
we will also need to add $\partial$ as an additional argument for the reconstructed 2-form,
$B_{\mu\nu}(\{B_{ab}\}, \osimp, \partial)$.
Note that, with $\epsilon$ now the only remaining background structure,
$B_{\mu\nu}(\{B_{ab}\}, \osimp, \partial)$ is now covariant with respect to
\textit{all orientation-preserving diffeomorphisms}.
Let $\partial$ and $\partial'$ be any two flat connections.
Any two such connections are related by an orientation-preserving diffeomorphism $\varphi$.\footnote{This
can be proven as follows. Because they are flat and $\mathcal{M} \cong \R^{3,1}$,
there exists coordinate systems $x^{\mathbf{a}}$ and $x^{\mathbf{a}'}$ such that
$\partial$ and $\partial'$ are the respective coordinate derivatives. Furthermore, because negating one coordinate in a
coordinate system does not change the associated coordinate derivative, these coordinate systems may be chosen without loss of generality to have the same orientation.  The unique diffeomorphism mapping  $x^{\mathbf{a}}$ to $x^{\mathbf{a}'}$
is then an orientation-preserving diffeomorphism mapping $\partial$ to $\partial'$.
}
Because $\varphi$ is orientation-preserving and maps $\partial$ to $\partial'$, if $\osimp$ is an ordered
4-simplex with respect to $(\partial, \epsilon)$, $\varphi \cdot \osimp$ is an ordered 4-simplex with respect to
$(\partial', \epsilon)$.  We therefore have
\begin{align*}
\omega(\{B_{ab}\}, \partial')
&:= \omega(B_{\mu\nu}(\{B_{ab}\}, \varphi\cdot \osimp, \partial'))
= \omega(B_{\mu\nu}(\{B_{ab}\}, \varphi \cdot \osimp, \varphi \cdot \partial))\\
& = \omega(\varphi\cdot B_{\mu\nu}(\{B_{ab}\}, \osimp, \partial))
= \omega(B_{\mu\nu}(\{B_{ab}\}, \osimp, \partial)) \\
&= \omega(\{B_{ab}\}, \partial)
\end{align*}
where the invariance of $\omega(B_{\mu\nu})$ under all orientation-preserving diffeomorphisms has been used.
This proves that $\omega(\{B_{ab}\}, \partial)$ is independent of the choice of $\partial$, and $\partial$ will be
dropped as an argument from now on.
Similarly, because $\nu(B_{\mu\nu})$ is also invariant under orientation preserving diffeomorphisms,
the exact same argument applies to $\nu$: $\nu(\{B_{ab}\}, \partial)$ is independent of $\partial$,
and this argument will be dropped from now on.

To see the independence of $\omega(\{B_{ab}\})$ and $\nu(\{B_{ab}\})$ from the choice
of fixed orientation of $\mathcal{M}$, let us now include the orientation $\epsilon$ explicitly as an argument,
$\omega(\{B_{ab}\}, \epsilon)$, $\nu(\{B_{ab}\}, \epsilon)$.
Because the definition of $\omega$ in terms of $B_{\mu\nu}$ also depends on $\epsilon$,
we furthermore write $\omega(B_{\mu\nu}, \epsilon)$.  We then have the following three facts:
(1.) For a given numbered 4-simplex $\geosimp$, flipping $\epsilon$ flips the orientation of the triangles $\Delta_{ab}$, and hence, via equation (\ref{disccont}) flips the sign of the reconstructed $B_{\mu\nu}$, so that
$B_{\mu\nu}(\{B_{ab}\}, \geosimp, -\epsilon) = - B_{\mu\nu}(\{B_{ab}\}, \geosimp, \epsilon)$.
(2.) When we flip $\epsilon$, the notion of which numbered 4-simplices are `ordered', and hence allowed in calculating
$\omega$ and $\nu$, is reversed.  Thus, if $\osimp$ is ordered with respect to $\epsilon$, and $P$ denotes parity,
then $P\osimp$ is ordered with respect to $-\epsilon$.
(3.) From equation (\ref{omegadef}), $\omega(B_{\mu\nu}, -\epsilon) = - \omega(B_{\mu\nu}, \epsilon)$.
Using these three facts, we have
\begin{align*}
\omega(\{B_{ab}\}, - \epsilon)
&:= \omega(B_{\mu\nu}(\{B_{ab}\}, P\osimp, -\epsilon), -\epsilon)
= - \omega(B_{\mu\nu}(\{B_{ab}\}, P\osimp, -\epsilon), \epsilon) \\
& = - \omega(- B_{\mu\nu}(\{B_{ab}\}, P\osimp, \epsilon), \epsilon)
= - \omega(P^* B_{\mu\nu}(\{B_{ab}\},\osimp, \epsilon), \epsilon)  \\
& = \omega(B_{\mu\nu}(\{B_{ab}\},\osimp, \epsilon), \epsilon)
= \omega(\{B_{ab}\}, \epsilon)
\end{align*}
and
\begin{align*}
\nu(\{B_{ab}\}, - \epsilon)
&:= \nu(B_{\mu\nu}(\{B_{ab}\}, P\osimp, -\epsilon))
= \nu(- B_{\mu\nu}(\{B_{ab}\}, P\osimp, \epsilon))
= \nu(P^* B_{\mu\nu}(\{B_{ab}\},\osimp, \epsilon)) \\
& = \nu(B_{\mu\nu}(\{B_{ab}\},\osimp, \epsilon))
= \nu(\{B_{ab}\}, \epsilon)
\end{align*}
where $\osimp$ is any 4-simplex ordered with respect to $\epsilon$.
Thus $\omega$ and $\nu$ for the reconstructed 2-form $B_{\mu\nu}$ do not
even depend on the background choice $\epsilon$, but depend on the algebra elements $\{B_{ab}\}$
alone, so that one can truly write $\omega(\{B_{ab}\})$ and $\nu(\{B_{ab}\})$.
%

For the following theorem we now fix an arbitrary isomorphism $\mathring{e}: \mathcal{M} \rightarrow \R^{3,1}$
preserving orientation and affine structure.
Let $\mathring{e}^I_\alpha$ denote the associated push-forward map; this then forms a tetrad on $\mathcal{M}$ which is
constant with respect to $\partial$.
Let $\mathring{g}_{\alpha\beta}=\mathring{e}_{\alpha}^I\mathring{e}_{\beta I}$ denote the associated metric,
and $\orM_{\alpha\beta\gamma\delta}=
\mathring{e}^0_{\alpha}\wedge\mathring{e}^1_{\beta}\wedge\mathring{e}^2_{\gamma}\wedge\mathring{e}^3_{\delta}$
the associated oriented volume form.  Using this metric structure, we let $N_a^\alpha$ denote the outward unit normal
to tetrahedron $\tau_a$ of $\geosimp$, and for each triangle $\Delta_{ab}$
we let $\mathring{\epsilon}^{\alpha\beta}_{[ab]}$ denote the associated (inverse) area form oriented as in definition
\ref{orientedTriangle}. It follows that
\begin{align*}
\orM^{\alpha\beta}_{[ab]}= \lambda \epsilon^{\alpha\beta}_{[ab]} =
\lambda\, \orM^{\alpha\beta\gamma\delta} (N_a)_{\gamma} (N_b)_{\delta},
\end{align*}
where the positive function $\lambda$ is fixed by the equation $2 = \mathring{g}_{\alpha\gamma} \mathring{g}_{\beta\delta}\orM^{\alpha\beta}_{[ab]}\orM^{\gamma\delta}_{[ab]}$.  One obtains, as in the Euclidean
case
\begin{equation*}
\lambda=|N_a\wedge N_b|^{-1}~.
\end{equation*}
Furthermore, define $N_a^I:= \mathring{e}^I_\alpha N_a^\alpha$ and
$B_{ab}^{\geom}(\geosimp):= B_{ab}^\geom(\mathring{e}(\geosimp))$.  We then have
\begin{theorem}
\label{geomform}
For any \emph{numbered} 4-simplex $\geosimp$,
\begin{align}
\label{Bform}
B_{\mu\nu}^{IJ}(\{B_{ab}^{\geom}(\geosimp)\},\geosimp) =
(\ast \mathring{e} \wedge \mathring{e})^{IJ}_{\mu\nu}
\end{align}
from which follows
\begin{align}
\label{nuone}
\nu(B_{\mu\nu}(\{B_{ab}^{\geom}(\geosimp)\},\geosimp)) = 1
\end{align}
and
\begin{align}
\label{omegaone}
\omega(B_{\mu\nu}(\{B_{ab}^{\geom}(\geosimp)\},\geosimp)) = 1
\end{align}
\end{theorem}
{\startproof
The first assertion was already proven in {\jonSec}. In Particular one computes
\begin{align*}
\int_{\Delta_{ab}} \mathring{e}^I\wedge \mathring{e}^J
= \int_{\Delta_{ab}} \orM^{[ab]}\;\orM^{\mu\nu}_{[ab]}\; \mathring{e}^I_{\mu}\; \mathring{e}^J_{\nu}
=\frac{1}{2}\,\epsilon^{IJ}{}_{KL} \;A(\Delta_{ab})\,\frac{(N_a\wedge N_b)^{KL}}{|N_a\wedge N_b|}
= A(\Delta_{ab})\frac{\left(\ast N_a \wedge N_b\right)^{IJ}}{|N_a \wedge N_b|}~.
\end{align*}
where $\int_{\Delta_{ab}} \mathring{\epsilon}^{[ab]} = A(\Delta_{ab})$ has been used.
Since $\ast^2 = -1$, using (\ref{eqn:geomB}) it follows that
\begin{align}
\label{eqn:contB}
B_{ab}^{\geom}(\geosimp)^{IJ}=\int_{\Delta_{ab}(\geosimp)}
(\ast \mathring{e}\wedge\mathring{e})^{IJ}
\end{align}
implying equation (\ref{Bform})\footnote{Note
the importance of the sign convention in (\ref{eqn:geomB}) for reaching this conclusion.
}, which immediately implies equation (\ref{nuone}).  Equation (\ref{omegaone})
then follows by
$\omega(B_{\mu\nu}(\{B_{ab}\},\geosimp)) :=
\mathrm{sgn}(\epsilon^{\alpha\beta\gamma\delta} \mathring{\epsilon}_{\alpha\beta\gamma\delta}) = 1$.
\finishproof}
Note this theorem applies independent of whether $\geosimp$ is ordered.
If the numbered 4-simplex $\geosimp$ is ordered then by definition
\bq
\omega(\{B_{ab}^{\geom}(\geosimp)\})\cdot\nu(\{B_{ab}^{\geom}(\geosimp)\}):=\omega(B_{\mu\nu}(\{B_{ab}^{\geom}(\geosimp)\},\geosimp))\cdot\nu(B_{\mu\nu}(\{B_{ab}^{\geom}(\geosimp)\},\geosimp))=1
\eq
where the last equality follows from the above theorem. If on the other hand $\geosimp$ is not ordered then $P\geosimp$ must be ordered, so that, using relations \eqref{eqn:omegaP}
\begin{align*}
\omega(\{B_{ab}^{\geom}(\geosimp)\})\cdot\nu(\{B_{ab}^{\geom}(\geosimp)\})&:=\omega(B_{\mu\nu}(\{B_{ab}^{\geom}(P\geosimp)\},\geosimp))\cdot\nu(B_{\mu\nu}(\{B_{ab}^{\geom}(P\geosimp)\},\geosimp)) \\
&=
\omega(B_{\mu\nu}(\{B_{ab}^{\geom}(\geosimp)\},\geosimp))\cdot\nu(B_{\mu\nu}(\{B_{ab}^{\geom}(\geosimp)\},\geosimp))
=1.
\end{align*}
Thus we have
\begin{lemma}
\label{lem:help}
For any numbered 4-simplex $\geosimp$, $\omega(\{B_{ab}^{\geom}(\geosimp)\})\nu(\{B_{ab}^{\geom}(\geosimp)\})=1$.
\end{lemma}
This lemma is the last piece needed to prove an extended version of the reconstruction theorem:
\begin{theorem}
\label{sectresult}
Suppose $\{A_{ab}, \bm{n}_{ab}\}$ is a set of non-degenerate boundary data and $\{X_a\}$ a solution of the orientation constraint $B_{ab}=-B_{ba}$ for $B_{ab}:=-A_{ab}\; \widehat{X}_a\triangleright \left[\timeo\wedge(0,\bm{n}_{ab})\right]$.
\begin{enumerate}
\item[(i)] If $\{X_a\}$ is not equivalent to a subset in $\SU(2)$, then $\{B_{ab}\}$ is either in Plebanski sector (II+) or (II-). Furthermore, there exists a numbered 4-simplex $\geosimp$ such that
$B_{ab}=\omega\nu B_{ab}^{\geom}(\geosimp)$.
\item[(ii)]If $\{X_a\}$ is equivalent to a subset in $\SU(2)$, then $\{B_{ab}\}$ is in the degenerate Plebanski sector.
\end{enumerate}
\end{theorem}
{\startproof

\noindent\textit{Proof of (i):}
From lemma \ref{lem:non-deg} and theorem \ref{th:bivector_geometry} follows the existence of a numbered 4-simplex $\geosimp$ such that  $B_{ab}=\mu\; B_{ab}^{\geom}(\geosimp)$. By lemma \ref{lem:help}:
\bq
\omega(\{B_{ab}\})\; \nu(\{B_{ab}\})
= \omega(\{\mu B^{\geom}_{ab}(\geosimp)\})\; \nu(\{\mu B^{\geom}_{ab}(\geosimp)\})=
\omega(B^{\geom}_{ab}(\geosimp))\; \nu(B^{\geom}_{ab}(\geosimp))\;\mu=\mu
\eq
which in addition implies $\nu(\{B_{ab}\}) \neq 0$, so that $\{B_{ab}\}$ is either in Plebanski sector (II$+$) or (II$-$).

\noindent\textit{Proof of (ii):}
If $\{X_a\}$ is equivalent to a subset in $\SU(2)$ then lemma \ref{lem:non-deg} implies that the bivectors $\{B_{ab}\}$ span a 3-dimensional hypersurface orthogonal to some fixed $N^I$.
Therefore $B_{\mu\nu}^{IJ}(\{B_{ab}\},\geosimp)$ associated to an arbitrary numbered 4-simplex $\geosimp$ must obey
\begin{displaymath}
0 = \left(\int_{\Delta_{ab}} (\ast B)^{IJ}\right) N_J
= \int_{\Delta_{ab}} \left((\ast B)^{IJ} N_J\right)
\end{displaymath}
for all $a, b$. Because $(\ast B)^{IJ} N_J$ is constant, it follows that
$(\ast B)^{IJ} N_J \equiv 0$,  from which one can show
\begin{displaymath}
\eta^{\mu\nu\rho\sigma} \epsilon_{IJKL} B_{\mu\nu}^{IJ} B_{\rho\sigma}^{KL}
=
\eta^{\mu\nu\rho\sigma} \epsilon_{IJKL} (\dual B)_{\mu\nu}^{IJ}
(\dual B)_{\rho\sigma}^{KL} = 0
\end{displaymath}
so that $B_{\mu\nu}^{IJ}(\{B_{ab}\},\geosimp)$ is in the degenerate sector for any $\geosimp$.
\finishproof}

\subsection{Restriction to the Einstein-Hilbert sector}
\label{sec:Einstein-Hilbert}

In order to permit a correct classical limit and have a clean relation between the spin foam model
and the canonical theory, it is necessary to restrict the solutions of the boundary value problem for 4-simplexes to the sector $\mu=1$. It is precisely in this sector that the BF action reduces to the Einstein-Hilbert action, and therefore it will be called
\emph{Einstein-Hilbert sector}. To later turn $\mu=1$ into an operator constraint it is vital to rephrase it in terms of the boundary data and the variables $X_a$.

Here and throughout the remaining part of the paper, $\hateq$ denotes equality modulo multiplication by a positive real number.
\begin{lemma}
Suppose $\{A_{ab},\bm{n}_{ab}\}$ is a set of Regge-like boundary data and $\{X_a\}\subset\SL$ a solution of the orientation constraint giving rise to a non-degenerate bivector geometry $\{B_{ab}\}$. Then there exist a numbered 4-simplex $\geosimp$ such that
\bq
B_{ab}^{\geom}(\geosimp)\hateq\beta_{ab}(\{\widehat{X}_{a'b'}\})\;(\widehat{X}_a\timeo)\wedge (\widehat{X}_b\timeo)
\eq
with $X_{ab}=X_{a}^{-1}X_{b}$ and
\bq
\beta_{ab}(\{\widehat{X}_{a'b'}\}):= \mathrm{sgn}\left[\epsilon_{ijk}(\widehat{X}_{ac}\timeo)^i(\widehat{X}_{ad}\timeo)^j(\widehat{X}_{ae}\timeo)^k\; \epsilon_{lmn}(\widehat{X}_{bc}\timeo)^l(\widehat{X}_{bd}\timeo)^m(\widehat{X}_{be}\timeo)^n\right]
\eq
where $\{c,d,e\}=\{0,\dots,4\}\setminus\{a,b\}$ in any order.
\end{lemma}
{\startproof

Recall from (\ref{Neq}) that the outward normal $N_a$ of the tetrahedron $\tau_a$ equals $\widehat{X}_a\timeo$ up to a sign $\epsilon_a$, so that from (\ref{eqn:geomB}) one has
$B_{ab}^{\geom}(\geosimp)\hateq-\epsilon_a\,\epsilon_b\, (\widehat{X}_a\timeo)\wedge (\widehat{X}_b\timeo)$.
As in the Euclidean case, the Gauss law imposes four-dimensional closure, $\sum_a V_aN_a=0$ (see Appendix B of {\jonProp}), implying
\bq
N_a=- V_a^{-1}\sum\limits_{b:b\neq a}V_b\, N_b
\eq
if $V_a\neq 0$. For space-like non-degenerate tetrahedra the metric volume $V_a$ is strictly positive so that we can repeat the calculation in {\jonProp} to show
\begin{align*}
0&<\epsilon(N_a,N_c,N_d,N_e)^2= -\frac{V_b}{V_a} \epsilon(N_b,N_c,N_d,N_e)\;\epsilon(N_a,N_c,N_d,N_e) \\
& \hateq  -\epsilon_a \epsilon_b \epsilon(\widehat{X}_b \timeo,\widehat{X}_c \timeo,\widehat{X}_d \timeo,\widehat{X}_e\timeo)\;
\epsilon(\widehat{X}_a \timeo,\widehat{X}_c \timeo,\widehat{X}_d \timeo,\widehat{X}_e \timeo)
\hateq -\epsilon_a\,\epsilon_b\; \beta_{ab}(\{\widehat{X}_{a'b'}\})
\end{align*}
Thus $\beta_{ab}=-\epsilon_a \epsilon_b$, which proves the lemma.
\finishproof}
An explicit expression for $\mu$ in terms of $\{X_a\}$ and $\{\bm{n}_{ab}\}$ is then
\begin{align*}
\mu&\hateq B^{\geom}_{ab}(\geosimp)_{IJ} B_{ab}^{IJ}
\,\hateq \,
\beta_{ab}(\{\widehat{X}_{a'b'}\})\left[(\widehat{X}_a\timeo)\wedge (\widehat{X}_b\timeo)\right]_{IJ}
 \left[-\widehat{X}_a\triangleright \timeo\wedge(0,\bm{n}_{ab})\right]^{IJ}\\[4pt]
& =-\beta_{ab}(\{\widehat{X}_{a'b'}\})\left[\timeo\wedge \widehat{X}_{ab}\timeo\right]_{IJ}
 \left[ \timeo\wedge(0,\bm{n}_{ab})\right]^{IJ}
 =2 \, \beta_{ab}(\{\widehat{X}_{a'b'}\}) \, (\widehat{X}_{ab}\timeo)_i\, \bm{n}_{ab}^i~.
 \end{align*}
The minus in the last equality is canceled due to $\timeo^I \timeo_I = -1$.
By using the isomorphism $\rho$, one has the more explicit expression
\begin{align}
\label{eqn:projX}
(\widehat{X}_{ab}\timeo)^i= \frac{1}{2}\tr\left(\pauli^i\,X_{ab}\,X_{ab}^{\dagger}\right) ~,
\end{align}
so that $\mu\hateq  \beta_{ab}(\{\widehat{X}_{a'b'}\}) \, \tr\left(\pauli^i\,X_{ab}\,X_{ab}^{\dagger}\right) \, \bm{n}_{ab}^i$.
If on the other hand $\{X_a\}$ is equivalent to a subset in $\SU(2)$ then clearly
$X_{ab}\subset\SU(2)$ and thus $\tr\left(\pauli^i\,X_{ab}\,X_{ab}^{\dagger}\right)=\tr\, \pauli^i=0$. This shows the validity of
\begin{theorem}
\label{proporthm}
For a given set of non-degenerate boundary data $\{A_{ab},\bm{n}_{ab}\}$ and a set $\{X_a\}$ solving orientation, the resulting bivectors $B_{ab}:=-A_{ab}\widehat{X}_a\triangleright \timeo\wedge (0,\bm{n}_{ab})$ are in the Einstein-Hilbert sector iff
\begin{align}
\beta_{ab}(\{\widehat{X}_{a'b'}\}) \, \tr\left(\pauli^i\,X_{ab}\,X_{ab}^{\dagger}\right) \, \bm{n}_{ab}^i >0
\end{align}
for any given pair $(a,b)$.
\end{theorem}

\section{Quantization and asymptotics of the EPRL-vertex}
\label{eprlsect}

To recall how the above classical variables relate to the structure of the quantum theory and to clarify the notation, this section gives a brief review of the quantization and asymptotic expansion of the Lorentzian EPRL-model based on {\asymptoticE} and {\asymptoticL}.

\subsection{Phase space and kinematical quantization}
The first step in the quantization of the model is to replace the BF-part of the action \eqref{BFaction} by its discrete counter part (see {\LorentzEPRL}).  Let a triangulation $\mathfrak{T}$ of $\mathcal{M}$ be given, with triangles denoted by $\Delta$ and
tetrahedra denoted by $\tau$.
The discrete variables then consist in an element $b_\Delta(\tau) \in \mathfrak{sl}(2,\C)$
for each tetrahedron $\tau$ and triangle $\Delta$ therein, and an element
$X_{\geosimp\tau} \equiv X_{\tau \geosimp}^{-1} \in SL(2,\C)$ for each 4-simplex $\geosimp$ and tetradhedron $\tau$ therein.
The discrete action is then \cite{Engle:2007wy}:
%
%
\begin{align}
\label{eqn:discS}
S_{disc}:=
-\frac{1}{2\kappa}\!\sum_{\Delta \in\mathrm{Int}(\mathfrak{T})}\!\!\!\tr\left[\{b_\Delta(\tau)
+\tfrac{1}{\gamma}\! \ast \! b_\Delta(\tau)\} X_\Delta(\tau)\right]
-\frac{1}{2\kappa}\!\sum_{\Delta\in\partial\mathfrak{T}}\!\!\!\tr\left[\{b_\Delta(t_\Delta)
+\tfrac{1}{\gamma}\! \ast \! b_\Delta(\tau_\Delta)\} X_\Delta \right] .
\end{align}
Here, for $\Delta$ in the interior $\mathrm{Int}(\mathfrak{T})$,
$X_{\Delta}(\tau):= X_{\tau \geosimp_1} X_{\geosimp_1 \tau_1} X_{\tau_1 \geosimp_2} \cdots X_{\geosimp_n \tau}$
with the product going around the `link' of $\Delta$, starting at $\tau$ and going in the direction determined
by the orientation of $\Delta$. For $\Delta$ in the boundary $\partial\mathfrak{T}$,
$X_{\Delta}(\tau):= X_{\tau \geosimp_1} X_{\geosimp_1 \tau_1} X_{\tau_1 \geosimp_2} \cdots X_{\geosimp_n \tau'}$
with the product going around the `link' of $\Delta$, starting and ending within the boundary of $\mathfrak{T}$,
in the direction determined by the orientation of $\Delta$.  Furthermore, for $\Delta \in \partial\mathfrak{T}$, $\tau_\Delta$
denotes the tetrahedron `above' $\Delta$ within the boundary.
Each tetrahedron $\tau$ and each 4-simplex $\geosimp$ has its own `frame'.
The algebra element $b_\Delta(\tau)$ is in the frame of $\tau$, and
the group element $X_{\geosimp\tau}$ plays the role of the `parallel transport' from the frame of $\tau$
to the frame of $\geosimp$.
%
%
On a single numbered 4-simplex $\geosimp$ the action reduces to the boundary term only.
If we additionally impose the time gauge in each tetrahedron frame so that the normal to each $\tau$
in its own frame equals $\timeo$ then the variables $b_\Delta(\tau)$ and $X_{\geosimp\tau}$
become exactly the previously defined variables $b_{ab}$ and $X_{a}$, with identification
$X_a \equiv X_{\geosimp\tau_a}, b_{ab} \equiv b_{\Delta_{ab}}(\tau_a)$.
The boundary phase space is then parameterized by $b_{ab}$ and
$X_{ab}:= X_a^{-1} X_b$.

Due to the extra term $\frac{1}{\gamma}\ast\!b\, X$ in the action, it is however not the $b$-field that is canonically conjugate to $X$ but $J:=\frac{1}{\kappa}\left(b+\frac{1}{\gamma}\! \ast \!b\right)$
(see \cite{Engle:2007wy,Engle:2007mu}).
In other words, the Hamiltonian vector fields of $J$ are the right invariant vector fields on the group and the Poisson-algebra of the matrix elements $J^{IJ}$ is isomorphic to $\slc$. With the usual definition of rotation and boost  generators,
\begin{align*}
\rot^i=\frac{1}{2}\epsilon^i_{jk} J^{jk}
\text{ and }\boost^i=J^{0i},
\end{align*}
the linear simplicity constraint (\ref{gflinsimp})
becomes
\bq
J_{ab}^{0j}+\gamma \ast J^{0j}_{ab}=\boost^j_{ab}+\gamma \rot^j_{ab}\approx 0
\eq
so that on the constraint surface we have
\begin{align}
\label{nLrelation}
A_{ab} \bm{n}_{ab}^i =
-(b_{ab})^{0i} = \frac{\kappa \gamma}{\gamma^2+1}(\rot^i_{ab}-\gamma \boost^i_{ab})
\approx \gamma \kappa \rot^i_{ab},
\end{align}
where (\ref{eqn:b_ab}) was used in the first equality.

With $X_{ab}$ as configuration variables, the kinematical Hilbert space $\hilbert_{\partial\geosimp}$ is the $L^2$
space over the ten copies of $\SL$ associated to the ten triangles in the boundary.
This space is spanned by generalized $\SL$ spin network functions based on the graph whose nodes $v_a$ are dual to the tetrahedra $\tau_a$ and whose links $\ell_{ab}$ are dual to the triangle $\Delta_{ab}$.
Each generalized spin network is specified by choosing, for each triangle $\Delta_{ab}$,
an irreducible representation of $\SL$ in the principal series -- specified by a real number $p_{ab}$ and an
half integer $k_{ab}$ -- together with two states $\Psi_{ab}, \Psi_{ba}$ in the carrying space $\mathcal{H}_{(p_{ab}, k_{ab})}$
of the chosen representation.  Explicitly it is given by
\begin{align}
\label{sltwoc_boundary}
\Psi_{\{k_{ab}, p_{ab}, \Psi_{ab}\}}(\{X_{ab}\})
:= \prod_{a<b} \langle \Psi_{ab}, X_{ab} \Psi_{ba} \rangle
\end{align}
The generators, $\boost_{ab}$ and $\rot_{ab}$, are quantized to act on the state $\Psi_{ab}$
via the generators on $\mathcal{H}_{(p_{ab}, k_{ab})}$ associated to $i\pauli$ and $\pauli$ respectively.

Each irrep of $\SL$ decomposes into a direct sum of $SU(2)$-irreducibles:
\bq
\mathcal{H}_{(k,p)}\simeq \bigoplus\limits_{j\geq k}\mathcal{H}_j~.
\eq
In the EPRL models \modernmodel, the simplicity constraints are imposed weakly by a master constraint 
which restricts the irreps $(p_{ab},k_{ab})$ to those obeying $p_{ab}=\gamma k_{ab}$ and the states
$\Psi_{ab}$ in $\mathcal{H}_{(k_{ab},p_{ab})}$ to the lowest $\SU(2)$ irreducible $\mathcal{H}_{k_{ab}}$.
The resulting reduced boundary Hilbert space $\mathcal{H}_{\partial\geosimp}^{EPRL}$
is then isomorphic to the $\SU(2)$ boundary Hilbert space of LQG.
To see this, we recall that the latter space is spanned by generalized $\SU(2)$ spin networks $\psi_{\{k_{ab}, \psi_{ab}\}}$,
each of which is labelled by a choice of $SU(2)$ irrep $k_{ab}$ and two states $\psi_{ab}, \psi_{ba} \in \mathcal{H}_{k_{ab}}$
for each triangle $\Delta_{ab}$.  Let $\mathcal{I}: \mathcal{H}_k \rightarrow \mathcal{H}_{(k,\gamma k)}$
denote the embedding of $\mathcal{H}_k$ into the lowest $\SU(2)$ irreducible in the decomposition of $\mathcal{H}_{(k,\gamma k)}$ above\footnote{If,
instead of setting $p=\gamma k$, one leaves $p$ free then the EPRL amplitude is suppressed anyway in the semiclassical
limit unless $p=\tilde{\gamma} k$
for some universal $\tilde{k}$, see {\asymptoticL}.
}.
The $\SL$ constrained boundary state corresponding to a generalized $SU(2)$ spin network is then given by
\begin{equation*}
\Psi_{\{k_{ab}, \psi_{ab}\}}(\{X_{ab}\}) :=
\prod_{a<b} \langle \mathcal{I} \psi_{ab}, X_{ab} \mathcal{I} \psi_{ba} \rangle
\end{equation*}
and the EPRL amplitude for the $\SU(2)$ boundary state $\psi_{\{k_{ab},\psi_{ab}\}}$ is simply the evaluation of the BF vertex amplitude on the associated constrained $\SL$ boundary state. That is,
\begin{align}
\label{eprlFromBF}
A_{\geosimp}^{EPRL}(\psi_{\{k_{ab}, \psi_{ab}\}}) := A_{\geosimp}^{BF}(\Psi_{\{k_{ab}, \psi_{ab}\}})
=\int_{\SL^5}\prod_{a=0}^4\dif X_a \delta(X_0) \prod_{a< b}\langle \mathcal{I} \psi_{ab},X_{ab} \mathcal{I} \psi_{ba} \rangle
\end{align}
where $\dif X$ denotes the Haar measure, and the $\delta(X_0)$ is inserted in order to gauge-fix the fifth group element, to ensure finiteness (see \cite{Engle:2008ev}).
In the next section we will give an explicit expression for the amplitude 
for coherent boundary states.

\subsection{Coherent states and the EPRL-amplitude}

\subsubsection{Representation theory and coherent states}

The unitary irreducible representations of $\SL$ can be realized on the space of homogeneous polynomials over $\C^2$ which scale as
\bq
f(\lambda z_0,\lambda z_1)=\lambda^{-1+ip+k}\bar{\lambda}^{-1+ip-k}f(z_0,z_1)
\eq
and on which the group acts by its transpose,
\bq
(Xf)(z)=f(X^T z)~,
\eq
so that rotation and boost generators act by respectively
$\rotoperator^i f:=-i\left(\frac{\dif\;}{\dif t}\right)_{t=0} \left(\ee{ \frac{i t}{2} \pauli^i} f\right)$ and
$\boostoperator^i f:=-i\left(\frac{\dif\;}{\dif t}\right)_{t=0} \left(\ee{ \frac{t}{2} \pauli^i} f\right)$.
The Hilbert space $\hilbert_{(k,p)}$ is then the subspace of square integrable functions with respect to the standard invariant 2-form $\Omega_z$ on $\C^2/{0}$,
$\Omega_z = \frac{i}{2}(z_0 dz_1 - z_1 dz_0) \wedge (\overline{z}_0 d\overline{z}_1 - \overline{z}_1 d\overline{z}_0)$.
Due to the scaling behavior of $\bar{f}\, g\, \Omega_z$ one can  define the scalar product as an integral over $\CP^1$:
\begin{align}
\label{eqn:scalar}
(f,g):=\int_{\CP^1} \bar{f}g\,\Omega~.
\end{align}
Furthermore, we can construct an invariant bilinear form $\alpha:\hilbert_{(p,k)}\times\hilbert_{(p,k)}\to\C$ using the unitary isomorphism $\mathcal{A}:\irrsl\to\hilbert_{(-k,-p)}$ with $\mathcal{A}^2=(-)^{2k}$ and the anti-unitary isomorphism $\irrsl\to\hilbert_{(-k,-p)}$ given by complex conjugation (see {\asymptoticL} for details). More precisely, we set
$\alpha(f,g):=({\cal J}f,g)$ where ${\cal J}f:=\overline{\mathcal{A} f}$. The bilinear form $\alpha$ is $\SL$-invariant, since ${\cal J}$ commutes with the group action {\asymptoticL}, and (anti-)symmetric, i.e. $\alpha(f,g)=(-)^{2k}\alpha(g,f)$.

The $\SU(2)$ irreducibles can be realized in a similar manner on the space of homogeneous functions of degree $2k$,
which scale as
\bq
\phi(\lambda z_0,\lambda z_1)=\lambda^{2k}\phi(z_0,z_1)
\eq
and on which the group acts again by its transpose.
In terms of this realization, the embedding
$\mathcal{I}:\hilbert_k \rightarrow \hilbert_{k, \gamma k}$
introduced in the last subsection can be
chosen to be of the form
\begin{align*}
\phi\mapsto\mathcal{I}\phi(z)=\scal{z,z}^{-1+i\gamma k-k}\phi(z)
\end{align*}
where $\scal{z,w}$ denotes the usual scalar product on $\C^2$. Note that ${\cal I}$ commutes with the action of $\SU(2)$ but not with the action of $\SL$. Moreover, if the inner product $(\cdot,\cdot)_k$ in $\mathcal{H}_k$ 
is defined by 
\begin{align}
\label{eqn:scalk}
(\phi,\phi)_k=\frac{\pi}{2k+1} \overline{\phi}^{a_1\dots a_{2k}}\,\phi_{a_1\dots a_{2k}},
\end{align}
where the coefficients $\phi^{a_1\dots a_{2k}}$ are extracted from the expansion $\phi(z)=\phi^{a_1\dots a_{2k}}z_{a1}\cdots z_{a_{2k}}$, then $\mathcal{I}$ is isometric \asymptoticE.
\\[5pt]
Coherent states provide a useful tool to relate states in $\hilbert_k$ to the classical geometry of a sphere  (see appendix \ref{app:coherent} for more details). In the fundamental representation, \textit{every} spinor $z$ is a coherent state
associated to the unit 3-vector ${\bf n}_z$ determined by the map
\begin{gather}
\begin{gathered}
\label{eqn:spinortovector}
\Gamma:\C^2\to\mathbb{H}\\
z\mapsto z\otimes z^{\dagger}=:\frac{\|z\|^2}{2}(\mathrm{Id}+\mathbf{n}_z \cdot \pauli).
\end{gathered}
\end{gather}
In other words, $z$ satisfies the eigenvalue equation 
\begin{equation*}
(\textbf{n}_z \cdot \pauli) z = z.
\end{equation*}
In the following we will restrict ourself to unit spinors $\xi$. If $\xi$ is associated to ${\bf n}_{\xi}$ then $J\xi$, where
\begin{align*}
J:\spinor{\xi_0}{\xi_1}\mapsto\spinor{-\bar{\xi}_1}{\bar{\xi}_0}
\end{align*}
is the anti-linear structure map of $\SU(2)$ for which $J^2=-\mrix{1}$ and $\scal{Jz,Jw}=\scal{w,z}$, is associated  to the norm $-{\bf n}_{\xi}$ pointing in opposite direction.

Since all irreducible representations $k$ of $\SU(2)$ can be obtained by taking the $2k$-fold symmetrized tensor product of the fundamental representations these geometric spinors can be easily lifted to any irrep. More precisely, we define the \emph{coherent states} in $\hilbert_k$ corresponding to $\xi \in \mathbb{C}^2$ by
\begin{align}
\label{eqn:su2cs}
C^k_{\xi}(z):=\sqrt{\frac{d_k}{\pi}}\langle \overline{\xi}, z \rangle^{2k},
\end{align}
where $d_k:=2k+1$.
The corresponding states in $\hilbert_{p,k}$ are then
\begin{align}
\label{eqn:CS}
\mathcal{I}C^k_{\xi}(z)=\sqrt{\frac{d_k}{\pi}}\scal{z,z}^{-1+ip-k}\langle \overline{\xi},z\rangle^{2k}~.
\end{align}
These states correspond to the normals $\bm{n}_{\xi}$ in the sense that
\begin{align*}
\bm{n}^i_{\xi}\,\rotoperator_i \,C^k_{\xi}= k\, C^k_{\xi}
\quad\text{and}\quad
(C^k_{\xi}|\rotoperator^i|C^k_{\xi})= k \,\bm{n}^i_{\xi}~.
\end{align*}
Other valuable characteristics of coherent states \eqref{eqn:CS} are their over completeness and the relation
\begin{align}
\label{scalarCS}
(C^k_{\xi'},C^k_{\xi})_k=\scal{\xi',\xi}^{2k}
\end{align}
which follows directly from equation \eqref{eqn:scalk} --- see appendix \ref{app:coherent} for a proof.

\subsubsection{EPRL-Amplitude for coherent boundary states}
\label{sssec:eprlcoh}
Recall that the reduced boundary Hilbert space can be parametrized by $\SU(2)$ spin network functions,
each labeled by a choice, for each triangle, of spin $k_{ab}$, and two vectors,
$\psi_{ab}$ and $ \psi_{ba}$, in the corresponding irrep of $SU(2)$.  We now choose each of these vectors $\psi_{ab}$
to be a coherent state $C^{k_{ab}}_{\xi_{ab}}$ determined by some spinor $\xi_{ab}$ as above.
This gives rise to a (Livine-Speziale \cite{Livine:2007vk})
\textit{coherent boundary state}. Obviously it is labeled by the choice of spins $k_{ab}$ and spinors $\xi_{ab}$, which
we correspondingly call \textit{quantum boundary data},
and is peaked on the classical boundary data $\{A_{ab}, \mathbf{n}_{ab}\}$ defined by
\begin{align}
\label{eqn:peaked}
A_{ab}=\gamma \kappa k_{ab}, \qquad \mathbf{n}_{ab}= \mathbf{n}_{\xi_{ab}}.
%
%
\end{align}
Because of the predominant use of the spins $k_{ab}$ throughout the rest of this paper, the set
$\{k_{ab}, \bm{n}_{ab}\}$ will also be called the classical reduced boundary
data from now on.
Furthermore a set of quantum boundary data $\{k_{ab}, \xi_{ab}\}$ will be said to satisfy closure, satisfy tetrahedron
non-degeneracy, or be Regge-like if the corresponding classical boundary data $\{k_{ab}, \mathbf{n}_{ab}\}$
satisfies the same condition.

Specializing the expression (\ref{eprlFromBF}) for the EPRL vertex amplitude to such a boundary state gives
\begin{align}
\label{eqn:vertexampl}
A_v=(-1)^{\Xi}\int_{\SL^5}\delta(X_4)\prod_a\dif X_a\prod_{a<b}{\cal P}_{ab}
\end{align}
where $(-1)^\Xi$ depends on the order of the tetrahedra and can be evaluated by a graphical calculus (see {\asymptoticL}),
and where\footnote{The expression used here for ${\cal P}_{ab}$ is different from that used in {\asymptoticL}. See appendix \ref{convapp}
regarding this change in convention.}
\begin{align*}
{\cal P}_{ab}:=\alpha(X_a\mathcal{I}C_{ab}, X_b\mathcal{I}C_{ba})
\end{align*}
 is called the \textit{propagator} for the triangle $(ab)$, with $C_{ab}:= C_{\xi_{ab}}$.

Using \eqref{eqn:CS},
the amplitude can be re-expressed (see appendix \ref{convapp}) as
%
%
\begin{align}
\label{eqn:vertexampl2}
A_v=(-1)^{\Xi} \,\, c \!\! \int_{\SL^5}\delta(X_0)\prod_a\dif X_a \int_{(\CP^1)^{10}}\ee{S}\prod_{a<b} \, d_{k_{ab}}\,\Omega_{ab}
\end{align}
where $c:=\frac{(1+\gamma^2)^5}{[\pi (1-i\gamma)]^{10}}$,
$\Omega_{ab}:= \frac{\Omega_{z_{ab}}}{\langle Z_{ab}, Z_{ab}\rangle \langle Z_{ba}, Z_{ba}\rangle}$,
\begin{align}
\label{eqn:simplexaction}
S[\{k_{ab},\xi_{ab}\}; \{X_a, z_{ab}\}]=\sum_{a<b} k_{ab}\,\log \frac{\scal{Z_{ab},\xi_{ab}}^2\scal{J\xi_{ba},Z_{ba}}^2}{\scal{Z_{ab},Z_{ab}}\scal{Z_{ba},Z_{ba}}}
+i\gamma k_{ab}\,\log \frac{\scal{Z_{ba},Z_{ba}}}{\scal{Z_{ab},Z_{ab}}},
\end{align}
and where we used the abbreviation $Z_{ab}:=X^{-1}_a z_{ab}$ and $Z_{ba}:=X^{-1}_b z_{ab}$.
The integration over the $z$-variables is due to the evaluation of $\alpha$ in each propagator.
Here we have written the argument of $S$ so that the boundary data is given first, followed by the integration variables.

\subsection{Asymptotic analysis}
\label{eprl_asympt}
The behavior of the amplitude \eqref{eqn:vertexampl2} for large spins can be studied by stationary phase methods, that is, the amplitude for boundary data $\{\lambda k_{ab}, \xi_{ab}\}$ in the regime $\lambda\to\infty$ is dominated by the critical points for which $\Re S$ is maximal and $\delta S=0$ (stationarity). As stated in {\asymptoticL}
(and translated according to appendix \ref{convapp} by $X \mapsto (X^\dagger)^{-1}$), the maximality of the real part of the action imposes
\begin{align}
\label{eqn:ReS}
X_a \xi_{ab}=\frac{\|Z_{ba}\|}{\|Z_{ab}\|} \ee{i\theta_{ab}} X_b J\xi_{ba}~,
\end{align}
while stationarity of the $z$-variables implies
\begin{align}
\label{eqn:deltaz}
(X_a^\dagger)^{-1}\xi_{ab}=\frac{\|Z_{ab}\|}{\|Z_{ba}\|} \ee{i\theta_{ab}} (X_b^\dagger)^{-1}J\xi_{ba}
\end{align}
for some set of phases $\theta_{ab}\in[0,2\pi]$ and where $\|Z\|^2=\scal{Z,Z}$.
From the variation of the group elements one obtains the closure conditions,
\begin{align}
\label{eqn:quantumsimplicity}
\sum_{b:b\neq a}k_{ab}\bm{n}_{ab}=0~,
\end{align}
%
%
%
and from equations \eqref{eqn:b_ab}, \eqref{Bdef} and \eqref{eqn:peaked} one reconstructs the physical bivectors as
\begin{align}
\label{eqn:Brecon}
\frac{1}{\gamma \kappa}B_{ab}:=
-k_{ab} \widehat{X}_a \triangleright \timeo\wedge (0,\bm{n}_{ab}) =
2k_{ab} \widehat{X}_a \triangleright \iota(\xi_{ab})\wedge\iota(J\xi_{ab}).
\end{align}
If the quantum boundary data $\{\xi_{ab},k_{ab}\}$ satisfy  linear simplicity, closure,
and tetrahedron non-degeneracy, then so do the above bivectors.  The
critical point equations \eqref{eqn:ReS} and \eqref{eqn:deltaz} then imply that furthermore orientation \eqref{orient}
is satisfied, so that the  bivectors $B_{ab}$ satisfy all weak bivector geometry conditions.
As we saw, these conditions in turn are sufficient to ensure the existence of a corresponding
constant, Plebanski 2-form which turns out to be simple --- and when 4-simplex non-degeneracy is additionally satisfied,
one knows that this 2-form is non-degenerate, and hence determines a space-time geometry.
%
%

All that remains is to evaluate the action at the critical points.
For this purpose it is convenient to fix a phase convention in the family of coherent boundary states considered.  Given classical
boundary data $\{k_{ab}, \bm{n}_{ab}\}$, the spinors $\xi_{ab}$ are, a priori, determined only up to a phase.
However, if the boundary data is Regge-like there exist a preferred geometrical choice such that
\begin{align*}
g_{ab}\xi_{ab}=J\xi_{ba}~.
\end{align*}
where $\{g_{ab}\} \subset \SU(2)$ are the quantum analogues of the gluing maps \eqref{eqn:gluing}.
States obeying this convention are called \emph{Regge-states}.
For all other types of boundary data there is no similarly natural convention so that the phase is left
arbitrary.

The amplitude of a Regge-state at a critical point depends only on $k_{ab}$ and the dihedral angle $\Theta_{ab}$ for which $\mathrm{cosh}\,\Theta_{ab}=|N_a\cdot N_b|$ if the boundary glues to a Lorentzian 4-simplex or $\cos\Theta^E_{ab}=|N^E_a\cdot N^E_b|$ if it glues to a Euclidean 4-simplex with
normals $N_a^E$.
We quote the result from {\asymptoticL}:
\begin{theorem}[EPRL asymptotics]
\label{asym_thm}
Let $\mathcal{D}=\{\lambda k_{ab},\bm{n}_{ab}\}$
%
%
be a set of non-degenerate boundary data satisfying closure.
\begin{enumerate}
\item
If $\mathcal{D}$ is Regge-like and determines the boundary geometry of a Lorentzian 4-simplex, then in the limit $\lambda \rightarrow \infty$,
%
%
\begin{align}
\label{eqn:asymL}
 A_v\sim\left(\frac{1}{\lambda}\right)^{12}\left[N_+ \exp\left(i\lambda\gamma \sum_{a<b} k_{ab}\Theta_{ab}\right)
+
N_- \exp\left(- i \lambda\gamma\sum_{a<b} k_{ab} \Theta_{ab}\right)
\right]~.
\end{align}
\item
If $\mathcal{D}$ is Regge-like and determines the boundary geometry of an Euclidean 4-simplex, then in the limit $\lambda \rightarrow \infty$,
\begin{align}
\label{eqn:asymE}
 A_v\sim\left(\frac{1}{\lambda}\right)^{12}\left[N_+^E \exp\left(i\lambda \sum_{a<b} k_{ab}\Theta^E_{ab}\right)
+
N_-^E \exp\left(- i \lambda\sum_{a<b} k_{ab} \Theta^E_{ab}\right)
\right]~.
\end{align}
\item
If $\mathcal{D}$ forms a vector geometry not in the above cases, then
\begin{align}
\label{vec_asym}
A_v \sim \left(\frac{2\pi}{\lambda}\right)^{12} N
\end{align}
 in the limit $\lambda \rightarrow \infty$.
\item
If $\mathcal{D}$ is not a vector geometry, then
$A_v$ falls off faster than any inverse power of $\lambda$.
\end{enumerate}
The factors $N_+$, $N_-$, $N_+^E$, $N_-^E$ and $N$  are independent of $\lambda$ and given in {\asymptoticL}
\end{theorem}

From the discussion at the end of section \ref{ssec:disc}, only in the case of \eqref{eqn:asymL}
do the reconstructed bivectors, and hence the reconstructed continuum Plebanksi 2-form correspond
to a non-degenerate Lorentzian geometry.
Moreover, since the sign in the two terms is dictated by $\mu$ (see {\asymptoticL})
the only part of the amplitude which corresponds to the Einstein-Hilbert sector is the first summand of \eqref{eqn:asymL} (see section \ref{sec:Einstein-Hilbert}).

\section{A proposed proper vertex amplitude}
\label{sec:proper}

\subsection{Definition}

One would like the semiclassical limit of a spin foam to be dominated by the exponential of $i$ times the Regge action
$S_R= \frac{1}{\kappa}\sum_{a<b} A_{ab} \Theta_{ab}= \gamma \sum_{a<b} k_{ab}\Theta_{ab}$ and not by the `cosine' \eqref{eqn:asymL} since, heuristically, the amplitude is a discretized version of the path integral $\int{\cal D} e\, {\cal D}A\, e^{iS[e,A]}$ where $S[e,A]$ is the Plebanski action.
More decisively, when multiple 4-simplices are considered, the presence of two terms as in (\ref{eqn:asymL}) for \textit{each}
4-simplex leads to unphysical equations of motion dominating in the semiclassical limit \cite{Engle:2012yg},
%
%
and thus are possibly the root cause of the unphysical curvature constraints found in \cite{Hellmann:2012kz},
and may even be the root cause of certain divergences in spin foam sums \cite{clrrr2012}.
The previous analysis suggests that the reason for the appearance of the undesired terms lies in summing over all Plebanski sectors instead of only taking the Einstein-Hilbert sector into account.

As derived in section \ref{sec:Einstein-Hilbert}, the data $\{k_{ab},\bm{n}_{ab},X_a\}$ determine a
Plebanski 2-form in the Einstein-Hilbert sector iff
$\beta_{ab}(\{\widehat{X}_{a'b'}\}) \, \tr\left(\pauli^i\,X_{ab}\,X_{ab}^{\dagger}\right) \, \bm{n}_{ab}^i >0$.
Hence one would like to insert the quantity
\begin{align*}
\Theta\left[\beta_{ab}(\{\widehat{X}_{a'b'}\}) \, \tr\left(\pauli^i\,X_{ab}\,X_{ab}^{\dagger}\right) \, \bm{n}_{ab}^i \right]
\end{align*}
into the path integral, where $\Theta[x]:= 1$ if $x>0$and $0$ otherwise.
Since, from (\ref{nLrelation}), on the reduced boundary phase space $\bm{n}^i_{ab}\hateq \rot^i_{ab}$,
the above quantity can be replaced by the quantum operator
\begin{align}
\label{eqn:quantum_projector}
\Pi_{ab}(\{X_{a'b'}\}):=\Pi_{(0,\infty)}\left(\beta_{ab}(\{\widehat{X}_{a'b'}\})\;\tr(\pauli_i\, X_{ab}\,X_{ab}^{\dagger})\,
\rotoperator^i \right)
\end{align}
where $\Pi_{\cal S}(\hat{O})$ denotes the spectral projector onto the part ${\cal S}\subset\R$ of the spectrum of the operator $\hat{O}$. This yields a new amplitude
\begin{align}
\label{proper_vertex}
\Aprop(\{k_{ab}, \psi_{ab}\}):=(-1)^{\Xi}\int_{\SL^5}\delta(X_4)\prod_a\dif X_a\prod_{a<b}
\alpha(X_a\mathcal{I}\psi_{ab},X_b\mathcal{I}\,\Pi_{ba}\left(\{X_{ab}\}\right)\,\psi_{ba})
\end{align}
which we call the \emph{proper vertex amplitude}.

To ensure that the amplitude is well-defined one has to check that it is independent of the gauge fixing $\delta(X_4)$.
This will be proven below. Furthermore, as we shall see, the integrand is $\SU(2)$-invariant. 

It appears at the first sight that the positioning of $\Pi$ inside the edge propagators in (\ref{proper_vertex})
is somewhat arbitrary causing quantization ambiguities. However, as in the case of the Euclidean proper vertex,
it is possible to move the projector anywhere inside ${\cal P}_{ab}$ if properly transformed (see theorem \ref{th:arbit_pos} in appendix \ref{app:Properties}).

\subsection{Symmetries}
\label{ssec:symmetries}

As with the EPRL vertex, the proper vertex makes use of extra structure in its definition which is not $\SL$-invariant, and so seems to break $\SL$-invariance of the vertex: The choice of a particular normal $\mathcal{T}$ used to impose linear simplicity in each tetrahedron frame,
as well as (implicitly) the choice of a particular embedding $h$ of $SU(2)$ into the subgroup of $\SL$ preserving $\mathcal{T}$, which determines the embedding $\mathcal{I}: V_k \rightarrow V_{p,k}$ via the covariance relation $h(g) \circ \mathcal{I} = \mathcal{I} \circ g$ for all $g \in \SU(2)$.  However, these extra structures in fact do \textit{not} break the $\SL$-invariance
of the proper vertex amplitude for reasons we now show.

For each tetrahedon $a$, let an $\SL$-element $\Lambda_a$ be specified, and use this group element to transform
the extra structure
at the given tetrahedron.  One then obtains the following
manifestly $\SL$-covariant, transformed amplitude:
\begin{align*}
{}^{\{\Lambda_{a'}\}}\Aprop(\{k_{ab}, \psi_{ab}\})
:= \int_{\SL^5} \prod_a \dif X_a
\prod_{a<b} \alpha( {}^{\Lambda_a}\mathcal{I} \psi_{ab}, X_{ab} {}^{\{\Lambda_{a'}\}}\Pi_{ba}(\{X_{a'b'}\})
{}^{\Lambda_b}\mathcal{I} \psi_{ba})
\end{align*}
where ${}^{\Lambda_a}\mathcal{I}:= \Lambda_a \circ \mathcal{I}$ and
\begin{align}
\label{covproj}
{}^{\{\Lambda_{a'}\}}\Pi_{ba}(\{X_{a'b'}\}) := \Pi_{(0,\infty)}
({}^{\{\Lambda_{a'}\}}\beta_{ba}(\{X_{a'b'}\})\epsilon_{IJKL} (\widehat{\Lambda}_b \mathcal{T})^I
 (\widehat{X}_{ba} \widehat{\Lambda}_a \mathcal{T})^J \widehat{J}^{KL})
\end{align}
with
\begin{align*}
\dummy \hspace{-0.3cm}{}^{\{\Lambda_{a'}\}}\beta_{ba}(\{X_{a'b'}\})&:=
 \sgn\left[\epsilon_{IJKL} (\widehat{\Lambda}_a \mathcal{T})^I (\widehat{X}_{ac} \widehat{\Lambda}_c \mathcal{T})^J
(\widehat{X}_{ad} \widehat{\Lambda}_d \mathcal{T})^K  (\widehat{X}_{ae} \widehat{\Lambda}_e \mathcal{T})^L \right. \cdot \\
& \hspace{0.8cm} \left. \cdot \epsilon_{MNPQ} (\widehat{\Lambda}_b \mathcal{T})^M (\widehat{G}_{bc}\widehat{\Lambda}_c \mathcal{T})^N
(\widehat{X}_{bd}\widehat{\Lambda}_d \mathcal{T})^P  (\widehat{X}_{be}\widehat{\Lambda}_e \mathcal{T})^Q
\right]
\end{align*}

Using (\ref{eqn:projX}) and (\ref{eqn:discS}) it is easy to show that, for the case $\Lambda_a \equiv 1$, this expression reduces to the earlier expression (\ref{proper_vertex}) for the proper vertex.  The non-trivial result is that \textit{the above transformed vertex amplitude is in fact
independent of the elements $\Lambda_a$}.  The proof of this fact is formally identical to the corresponding proof given for the Euclidean proper vertex {\jonProp} with the replacements
$Spin(4) \mapsto \SL$, $SO(4) \mapsto SO(1,3)$,
$V_{j^-, j^+} \mapsto V_{p,k}$,
$\epsilon(\cdot,\cdot) \mapsto \alpha(\cdot, \cdot)$,
$\iota_k^{s^+ s^-} \mapsto \mathcal{I}$,  and with $h$ replaced with the inclusion
map $SU(2) \hookrightarrow \SL$. We thus do not repeat the proof here.

The above result implies, as an important corollary, the $SU(2)$ gauge-invariance of the proper vertex. Explicitly,
given any choice of $SU(2)$ gauge rotation $g_a$ at each tetrahedron, we have
\begin{align*}
\Aprop(\{k_{ab}, g_a \psi_{ab}\})
&= \int_{\SL^5} \prod_a \dif X_a \prod_{a<b}
\alpha( \mathcal{I} g_a \psi_{ab}, X_{ab}\Pi_{ba}(\{X_{a'b'}\}) \mathcal{I} g_b \psi_{ba}) \\
&= \int_{\SL^5} \prod_a \dif X_a \prod_{a<b}
\alpha( {}^{g_a}\mathcal{I} \psi_{ab}, X_{ab}\Pi_{ba}(\{X_{a'b'}\}) {}^{g_b} \mathcal{I} \psi_{ba}) \\
&= {}^{\{g_{a'}\}}\Aprop(\{k_{ab}, \psi_{ab}\})
= \Aprop(\{k_{ab}, \psi_{ab}\})
\end{align*}
where $\mathcal{I} g_a = g_a \mathcal{I} = {}^{g_a} \mathcal{I}$ as well as the
fact that (\ref{covproj}) reduces to ${}^{\{\Lambda_{a'}\}}\Pi_{ab} = \Pi_{ab}$
for $\{\Lambda_a\} \subset SU(2)$ were used.

\section{Conclusion}

We have shown in this paper that the Lorentzian EPRL model {\LorentzEPRL},
like the Euclidean model, 
mixes three of the Plebanski sectors and two dynamically determined orientations.
This is not completely surprising, since it is based, as the Euclidean model, 
on the linear simplicity constraint. 
We have furthermore shown why this mixing can be identified as the cause for the appearance of more than one
term in the asymptotic limit {\asymptoticL}. 
In order to cure this we here derived an additional constraint that isolates the sector in which the Plebanski action is equivalent to the Einstein-Hilbert action. The principal ideas within this derivation and also the final form of the constraint closely resemble those of the Euclidean theory {\jon}. However, since the splitting of bivectors into self- and anti-self-dual part is no longer available, we here had to rephrase everything in a language which more four dimensionally covariant. 
Because of this, the above derivation can be seen as a more fundamental one as it applies to both Euclidean and Lorentzian signatures. 
In the second part of this paper we derived a \emph{proper vertex amplitude} for the Lorentzian model, confirming the guess of {\jonProp}. The new amplitude is well-defined, linear in its boundary states, independent of the choice of background structure and $\SL$-invariant. 

Of course, in order to finally justify this modification, it is necessary to prove that the asymptotics is governed by only a single Regge term, which is technically involved because (a) the projector in the amplitude does not scale exponentially in the spins and 
(b) is discontinuous at points where $X^{\dagger}_{ab}=X_{ab}^{-1}$ for some $a<b$. In order to deal with these problems we will have to develop new tools going beyond the usual extended stationary phase method. This will be discussed in \cite{evz2015}. 
It would also be interesting to investigate whether the proper vertex can solve some of the issues in the asymptotic expansion on many simplices such as the appearance of non-gravitational terms \cite{Magliaro:2011dz,Han:2011re,Han:2011rf} and the appearance of an unphysical curvature constraint \cite{Hellmann:2012kz}. 

In addition, it is important to check whether the above modification affects the predictions of the model, for example the graviton propagator calculation \cite{Bianchi:2009ri, Alesci:2008ff} or spin foam cosmology \cite{Ashtekar:2009dn, Ashtekar:2010gz, Ashtekar:2010ve}. The graviton propagator for the present proper vertex amplitude is calculated in \cite{cev2015} where,
to lowest order in the vertex expansion,  the same answer as for the usual EPRL amplitude is found. 
If, however, more vertices are considered, we expect a deviation between the predictions of the standard EPRL model and the proper vertex, and suspect that only the proper vertex will deliver results consistent with linearized gravity.
At present this is just an expectation. 

A last, promising future research direction would be to pass over to the ``dual picture'' based on the coloring 
of the 2-complex dual to the triangulation by spins and intertwiners, 
and finally 
to extend the above derivation to arbitrary complexes and the KKL-model. This is a highly non-trivial task since the above derivation relies heavily on the combinatorics of four simplices in a number ways. 

{\acknowledgments
We thank Ilya Vilensky for discussions.  
J.E. was supported in part by NSF grants  PHY-1205968 and PHY-1505490,
and by NASA through the University of Central Florida's NASA-Florida Space Grant Consortium.
A.Z. acknowledges financial support of the `Elitenetzwerk Bayern' on the grounds of `Bayerische Elitef\"order Gesetz' and the support by the grant of Polish Narodowe Centrum Nauki
 nr 2012/05/E/ST2/03308.
}

\appendix
\section{Spinors and coherent states}
\label{app:coherent}

The coherent states used in this paper go back to Perelomov \cite{Perelomov:cs1972} and were first applied to spin foam models in \cite{Freidel:2007py, Livine:2007ya, Livine:2007vk}. For the sake of self-containedness some of the main properties of these states are reviewed. For further reading see e.g. \cite{PerelomovGCS}.

In the following the $SO(3)$ element associated to a given $SU(2)$ element $g$ will be denoted $\widehat{g}$.
Furthermore, we will denote the scalar product in $\R^3$ by a dot, the generators of rotation by $\rot^i:=\frac{1}{2}\pauli^i$ and its corresponding generator in the irreducible representation $(\hilbert_k,\rho_k)$ by
$\rotoperator_k^i:=-i\left(\frac{\dif\;}{\dif t}\right)_{t=0} \rho_k(\ee{ i t \rot^i})$.
%
%

\begin{definition}
\label{def:coherent}
Given $\bm{n}\in S^2$, we define the normalized state $\ket{\bm{n}; k,m}\in\hilbert_k$, $m\in\{-k,-k+1,\dots,k\}$ by
\begin{align}
\label{coheqns}
\bm{n}\cdot\rotoperator_k\; \ket{\bm{n}; k,m}=  m\, \ket{\bm{n}; k,m}\qquad \text{and} \qquad (\rotoperator_k)^2\ket{\bm{n}; j,m}=k(k+1)\ket{\bm{n}; k,m},
\end{align}
the phase of $\ket{\bm{n};k,m}$ being fixed arbitrarily.
\end{definition}
From (\ref{coheqns}) one proves that these states satisfy $\scal{\bm{n}; k,m| \rotoperator^i_k|\bm{n}; k,m}= m\,\bm{n}^i$.
\begin{lemma}
\label{lem:A}
Let $g\in\SU(2)$ then $\rho_k(g)\;\ket{\bm{n}; k,m}=\ee{i\phi_g}\;\ket{g\,\bm{n}; k,m}$ for some $\phi_g\in[0,2\pi]$.
\end{lemma}
{\startproof
The operator $\rotoperator$ transforms covariantly under $\SU(2)$ so that $g\triangleright\bm{n}\cdot\!\rotoperator=g\, \bm{n}\cdot\!\rotoperator g^{-1}=(g\,\bm{n})\cdot\!\rotoperator$. Thus both states, $\rho_k(g)\ket{\bm{n}; k,m}$ and $\ket{g\,\bm{n}; k,m}$ are eigenstates of
\bq
\left((g\,\bm{n})\cdot\!\rotoperator\right)_k=\rho_k(g)\left(\bm{n}\cdot\!\rotoperator\right)_k\rho_k(g^{-1})~.
\eq
\finishproof}
Thus, any state $\ket{\bm{n}; k,m}$ can be obtained from some reference state
$\ket{\bm{n}_o; k,m}$ by the action of an appropriate group element $g \in \SU(2)$ mapping $\bm{n}_o$ to
$\bm{n}$.\footnote{This is the original definition of a coherent state by Perelomov.}
When constructed in this way, normally one fixes $\bm{n}_0$ along the z-axis.

When one chooses $m=k$, the resulting states
$\ket{\bm{n};k}:= \ket{\bm{n};k,k}$ are \textit{coherent states}.  These are `coherent' in the sense that they minimize the uncertainty of the
generators of the group (see \cite{PerelomovGCS}) and therefore are best suited for semiclassical analysis. Furthermore, they have the advantage that $\ket{\bm{n};k}=\ket{\bm{n};1/2}^{\otimes 2k}$. This is because $\hilbert_k$ is isomorphic to the $2k$-fold symmetric tensor product of $\hilbert_{\frac{1}{2}}$ and
\begin{align}
\label{A_help1}
\rotoperator_k=\sum_{n=1}^{2k}\; \underbrace{1\otimes 1\otimes \cdots}_{(i-1)\text{ times}}\; \otimes \,\rotoperator_{1/2}\underbrace{\otimes\, 1\cdots\otimes 1}_{(n-i) \text{ times}}~.
\end{align}
In the following we will mostly use the states $\ket{\bm{n}; k}$.

As already mentioned in the main text, each unit spinor $\xi\in\C^2$, $||\xi||=1$, corresponds naturally to a normal with components
$\bm{n}_{\xi}^i=\tr\left(\xi\otimes\xi^{\dagger}\pauli^i\right)=2\scal{\xi|\rotoperator^i|\xi}$.
From $(\bm{n}_\xi \cdot\!\rotoperator)_{1/2}=\frac{1}{2}\bm{n}_\xi \cdot\pauli=\xi\otimes\xi^{\dagger}-\frac{1}{2} \mathbbm{1}$, it then directly follows that $\xi$ satisfies the conditions in definition \ref{def:coherent} for
$\bm{n}_\xi$, $k=m=1/2$, so that
$\xi = e^{i\theta} \ket{\bm{n}; \frac{1}{2}}$ for some phase $\theta$. In other words, every spinor $\xi$ is a coherent state in the fundamental representation. The contraction of two spinors yields the following geometrical expression:
\begin{align*}
|\scal{\xi',\xi}|^2=\tr\left[\Gamma(\xi')\,\Gamma(\xi)\right]=\frac{1+\bm{n}_{\xi'}\cdot\bm{n}_{\xi}}{2}~,
\end{align*}
where $\Gamma$ is the map \eqref{eqn:spinortovector}.

To generalize the above to arbitrary irreducible representations let us briefly review the realization of $\hilbert_k$ on the space $V_k$ of homogeneous polynomials of degree $2k$ over $\C^2$ on which the group acts by its transpose:
\begin{align}
\label{action}
(g f)(z)=f(g^T z)~.
\end{align}
The action of the generators can be computed as
\begin{align}
\label{A_help2}
i \rotoperator^i_k \,f(z)=\left(\frac{\dif\;}{\dif t}\right)_{t=0} f((\ee{t\frac{i}{2}\pauli^i})^T z)
= \left(\frac{\partial z_0(t)}{\partial t}\frac{\partial\;}{\partial z_0}+\frac{\partial z_1(t)}{\partial t}\frac{\partial\;}{\partial z_1}\right)_{t=0}\; f(z)
\end{align}
with $z(t)=(\ee{t\frac{i}{2}\pauli^i})^T z$. For example, the z-component is equal to
\begin{align}
\label{eqn:z-component}
\rotoperator^3_k=\frac{1}{2}\left(z_0\,\frac{\partial\;}{\partial z_0} -z_1\,\frac{\partial\;}{\partial z_1}\right)
\end{align}
for all $k$.
\\
Functions $f\in V_k$ can be expanded either in terms of monomials, i.e.
\bq
f(z)=\sum_{\mu=0}^{2k} f_k^{\mu} z_0^{\mu}  z_1^{2k-\mu}~,
\eq
or in terms of completely symmetric components $f^{a_1\dots a_{2k}}$ with $a_i= 0,1$, i.e.
\bq
f(z)=\sum_{a_1,\dots,a_{2k}} f^{a_1\dots a_{2k}} \,z_{a_1}\,\cdots\, z_{a_{2k}}~.
\eq
A comparison of both methods reveals $ f^{\mu}_k=\binom{2k}{\mu} f^{a_1\dots a_{2k}}$ where  $\sum_{i=1}^{2k} \,a_i=2k-\mu$. Moreover, the relation \eqref{action} implies that the coefficients $f^{a_1\dots a_{2k}}$ transform as elements of $\mathrm{Sym}[\C^2_1\otimes \cdots\otimes\C^2_{2k}]$. Explicitly,
\begin{align}
\label{componentaction}
(g\cdot f)^{a_1\dots a_{2k}}= g^{a_1}\,_{b_1} \;\cdots\,  g^{a_{2k}}\,_{b_{2k}}\,f^{b_1\dots b_{2k}}~.
 \end{align}
 \begin{lemma}
 The inner product on $V_k$ can be chosen such that
\begin{align}
\label{scalarproduct}
(f,h)_k:=\int_{\C^2}\dif\mu(z,k) \overline{f(z)}\, h(z)= \frac{\pi}{2k+1} \bar{f}^{a_1\dots a_{2k}}\, h_{a_1\dots a_{2k}}
\end{align}
where $\dif\mu(z,k)= \frac{1}{\pi (2k+1)!}\ee{-|z|^2} \dif z_0 \dif z_1$.
\end{lemma}
{\startproof
The integral can be easily computed using the polar coordinates $z_i=r_i\ee{i\theta_i}$ and the integral equalities
\bq \int_0^{2\pi} \dif \theta\ee{i\theta n}= 2\pi\,\delta_{n,0}\quad\text{and}\quad
\int_0^{\infty}\dif t\;t^n \ee{-t}=n!~.
\eq
We find:
\begin{align*}
(f,h)_k&=\frac{1}{\pi (2k+1)!} \int_0^\infty\dif r_0 \,\dif r_1\int_0^{2\pi} \dif \theta_0\,\dif \theta_1\,\ee{r_0^2+r^2_1}
 \sum_{\mu,\nu} \bar{f}^{\mu}_k\, h^{\nu}_k \,r_0^{\mu+\nu+1}\, r_1^{4k-\mu-\nu+1}\,\ee{i\theta_0(\nu-\mu)}\,\ee{i\theta_1(\mu-\nu)}\\
&=\frac{4\pi^2}{\pi (2k+1)!}  \sum_{\mu} \bar{f}^{\mu}_k h^{\mu}_k \int_0^\infty \dif r_0\,\dif r_1 \;r_0^{2 \mu+1} \,r_1^{4k-2\mu+1}\\
&= \frac{\pi}{(2k+1)!} \sum_{\mu} \bar{f}^{\mu}_k\, h^{\mu}_k\, \mu!\,(2k-\mu)!
= \frac{\pi}{2k+1} \bar{f}^{a_1\dots a_k} h_{a_1\dots a_k}~.
\end{align*}
The last equality follows from $ f^{\mu}_k=\binom{2k}{\mu} f^{a_1\dots a_{2k}}$ and the fact that there are exactly $\binom{2k}{\mu}$ combinations  $(a_1\dots a_{2k})$ summing up to $2k-\mu$.
From this final expression for $(f,h)_k$, it is clear that this inner product makes the action (\ref{componentaction})
of $SU(2)$ unitary, and so is an allowable choice.
\finishproof}
With respect to the measure of \eqref{scalarproduct} the normalized eigenstates of \eqref{eqn:z-component} with eigenvalue $m:=k-\mu$  can be chosen to be
\begin{align*}
f^k_m(z):=\left[\frac{\pi\;(k-m)! (k+m)!}{(2k+1)!}\right]^{-1/2} z_0^{k+m} z_1^{k-m}~.
\end{align*}
%
%
%
A coherent state in $V_k$ can be obtained by acting with the group element
\begin{align*}
g_{\xi}=\zweimatrix{\xi_0}{-\bar{\xi}_1}{\xi_1}{\bar{\xi}_0}
\end{align*}
on the canonical state $f^k_k(z)$. This leads to the normalized state
\begin{align}
\label{Ccohdef}
C^k_\xi(z):= \sqrt{\frac{2k+1}{\pi}}\scal{\bar{\xi},z}^{2k}=\sqrt{\frac{2k+1}{\pi}}\left[\xi^{\otimes 2k}\right]^{a_1\dots a_k}\, z_{a_1}\,\cdots \,z_{a_{2k}}~.
\end{align}
That this state meets the requirements of definition \ref{def:coherent} for the normal $\bm{n}_{\xi}$,
and hence is equal to $|\bm{n}_{\xi};k\rangle$ up to a phase, can be tested by either applying the derivative operators \eqref{A_help2} to the functions $C^k_\xi(z)$ or by acting with the algebraic counterpart \eqref{A_help1} on the components.
$\xi \mapsto C^k_\xi(z)$ is furthermore covariant under the action of any $g \in \SU(2)$:
\begin{align*}
C^k_{g\xi}(z) &=  \sqrt{\frac{2k+1}{\pi}}\scal{\overline{g\xi},z}^{2k}
=  \sqrt{\frac{2k+1}{\pi}}\scal{\bar{\xi},\bar{g}^\dagger z}^{2k}
=  \sqrt{\frac{2k+1}{\pi}}\scal{\bar{\xi},g^T z}^{2k}
=  C^k_{\xi}(g^T z)
= \left(gC^k_{\xi}\right)(z).
\end{align*}
From \eqref{scalarproduct}, one furthermore has
\begin{align}
\label{equal_scalar}
( C^k_{\xi'},C^k_{\xi})
=\frac{\pi}{2k+1}[\overline{ C^k_{\xi'}}]^{a_1\dots a_k}\,[C^k_{\xi}]_{a_1\dots a_k} = \scal{\xi',\xi}^{2k}~.
\end{align}

Another very important aspect of coherent states we have not mentioned so far is the over-completeness of the system $\{\ket{\bm{n}; k}\}_{\bm{n}\in S^2}$, in the sense that
\bq
B:=\int_{S^2}\dif\bm{n}\, \ket{\bm{n}; k}\bra{\bm{n}; k}=b_k\mrix{1}_k
\eq
where $\dif\bm{n}$ denotes the measure on the metric sphere with unit radius.
Note this is a direct consequence of lemma \ref{lem:A} and the irreducibility of $\hilbert_k$. Namely, lemma \ref{lem:A} and the invariance of the measure $\dif\bm{n}$ imply $\rho(g) B\rho(g^{-1})=B$ but since there are no invariant subspaces $B$ must be proportional to $\mrix{1}$. To determine $b_k$ we compute $\scal{\bm{n}|B|\bm{n}}$:
\begin{align}
\label{resolution1}
b_k&=\scal{\bm{n}|B|\bm{n}}=\int_{S^2}\dif\bm{m}\,|\scal{\bm{m},\bm{n}}|^2
=\int_{S^2}\dif\bm{m} \left(\frac{1+\bm{m}\cdot \bm{n}}{2}\right)^{2k}
=\frac{4\pi}{2k+1} .
\end{align}
Using the isomorphism
\begin{align}
\label{eqn:iota}
\iota: \CP^1 \rightarrow S^2, [\xi] \mapsto \bm{n}_\xi
\end{align}
the above resolution of the identity can be written in terms of the states $\{C^k_{\xi}\}$,
yielding a reproducing kernel
\begin{align}
\label{resolution2}
K(z,z'):=\frac{2k+1}{\pi}\int_{\CP^1}\Omega_{\hat{\xi}} \; C^k_{\hat{\xi}}(z)\; \overline{C^k_{\hat{\xi}}(z')}
\end{align}
in the space $V_k$ with measure $\dif\mu(z',k)$, so that
\begin{align}
\label{resolution3}
\int_{\C^2} \dif\mu(z',k)\; K(z,z')\, f(z')=
\frac{2k+1}{\pi}\int_{\CP^1}\Omega_{\hat{\xi}} \; C^k_{\hat{\xi}}(z) \;(C^k_{\hat{\xi}},\,f)_k
= f(z)~.
\end{align}
Here $\hat{\xi}:= \xi/||\xi||$, so that $\Omega_{\hat{\xi}} = \Omega_\xi/||\xi||^4$.
To see that the isomorphism $\iota$ indeed maps $\Omega_{\hat{\xi}}$ to $d\bm{n}/4$,
one can use the following inverse
\begin{align*}
\iota^{-1}: \bm{n}(\theta,\phi):= (\sin \theta \cos \phi, \sin \theta \sin \phi, \cos \theta) \mapsto \left[\xi = \left(1, e^{i\phi} \tan \frac{\theta}{2}\right)\right] .
\end{align*}
%
%
One then has
\begin{align*}
(\iota^{-1})^* \Omega_{\hat{\xi}}
= \frac{i}{2||\xi||^4}(\xi_0 d \xi_1 - \xi_1 d \xi_0)\wedge(\overline{\xi}_0 d \overline{\xi}_1
- \overline{\xi}_1 d \overline{\xi}_0)
= \frac{1}{4} \sin \theta d \theta \wedge d \phi .
\end{align*}
%
%

\section{Expression for the Lorentzian vertex with projectors on the left}
\label{app:Properties}

Most of the properties discussed in appendix E and D of {\jonProp} can be directly generalized to Lorentzian signature with minor modification. However, since these features are crucial for the analysis of the new vertex amplitude we will briefly review these results and adapt them to Lorentzian signature.

\begin{lemma}
\label{lemma9}
For any two irreducible unitary representations $(V,\rho_V)$ and $(W,\rho_W)$ of $\SU(2)$ and any $\SU(2)$-covariant,
isometric map $\mathfrak{H}:V\to W$, one has:
\begin{itemize}
\item[(a.)] $\rotoperator^i\; \mathfrak{H}= \mathfrak{H}\; \rotoperator^i$
\item[(b.)] $\mathfrak{H}\,\ket{\bm{n}; k,m} = e^{i\theta} \ket{\bm{n}; k,m}$
for some $\theta$.
\item[(c.)] $\Pi_{\cal S}(\bm{n}\cdot\!\rotoperator)\,\mathfrak{H}=\mathfrak{H}\;\Pi_{\cal S}(\bm{n}\cdot\!\rotoperator)$
\end{itemize}
\end{lemma}
{\startproof

\noindent{\it Proof of (a.):}
As before, $ \rotoperator^i:=-i\left(\frac{\dif\;}{\dif t}\right)_{t=0}\;\rho (\ee{\frac{i}{2}\pauli^i})$
are the generators in a given representation $\rho$. The result then follows from the fact that
$\mathfrak{H}$ commutes with the group action.
%
%
%

\noindent{\it Proof of (b.):} By definition \ref{def:coherent} $\ket{\bm{n}; k,m}$ is an eigenstate of $\bm{n}\cdot\!\rotoperator$ and $\rotoperator^2$ with eigenvalues $m$ and $k(k+1)$. By (a.), so is $\mathfrak{H}\,\ket{\bm{n}; k,m}$. \\
{\it Proof of (c.):} This is immediate from (a).

\finishproof}
\begin{lemma}
\label{lemma10}
For any $v^I, \timeo^I  \in\R^{3,1}$ and any $\Lambda\in\SL$:
\bq
\rho(\Lambda) \Pi_\mathcal{S}(\epsilon_{IJKL}\mathcal{T}^I n^J \hat{J}^{KL})=\Pi_\mathcal{S}\left(\epsilon_{IJKL}(\Lambda \mathcal{T})^I (\Lambda n)^J \hat{J}^{KL}\right) \rho(\Lambda)
\eq
\end{lemma}
{\startproof

Suppose $|\lambda\rangle \in \hilbert_{k,p}$ is an eigenstate of $\epsilon_{IJKL}\mathcal{T}^I n^J \hat{J}^{KL}$ with eigenvalue $\lambda$.
Then
\begin{align*}
\epsilon_{IJKL}(\Lambda \mathcal{T})^I (\Lambda n)^J \hat{J}^{KL} \rho(\Lambda) |\lambda \rangle
&= \rho(\Lambda) \epsilon_{IJKL}(\Lambda \mathcal{T})^I (\Lambda n)^J \Lambda^K{}_M \Lambda^L{}_N \hat{J}^{MN}
|\lambda \rangle \\
&= \rho(\Lambda) \epsilon_{IJKL} \mathcal{T}^I n^J \hat{J}^{KL}
|\lambda \rangle = \lambda \rho(\Lambda) |\lambda\rangle
\end{align*}
so that $\rho(\Lambda) |\lambda\rangle$ is an eigenstate of
$\epsilon_{IJKL}(\Lambda \mathcal{T})^I (\Lambda n)^J \hat{J}^{KL}$ with eigenvalue $\lambda$, and in the first equality
the covariance of $\hat{J}^{KL}$ was used.
We then have
\begin{align*}
\rho(\Lambda) \Pi_\mathcal{S}(\epsilon_{IJKL}\mathcal{T}^I n^J \hat{J}^{KL}) |\lambda \rangle
= \chi_\mathcal{S}(\lambda) \rho(\Lambda) |\lambda \rangle
= \Pi_\mathcal{S}\left(\epsilon_{IJKL}(\Lambda \mathcal{T})^I (\Lambda n)^J \hat{J}^{KL}\right) \rho(\Lambda) |\lambda \rangle.
\end{align*}
Because such states $|\lambda \rangle$ span $\hilbert_{k,p}$, the result follows.
\finishproof}

Like for the Euclidean proper amplitude, the projector can be positioned anywhere inside the edge propagators ${\cal P}_{ab}$ with appropriate transformation. To show this we still need to determine the effect of the anti-linear structure map ${\cal J}$
on the projectors.
\begin{lemma}
\label{lemma13}
For any ${\cal S}\subset\R$ and any $\bm{n}\in\R^3$,
\begin{align*}
\Pi_{\cal S}(\bm{n}\cdot \rotoperator)\, {\cal J}= {\cal J}\,\Pi_{\cal S}(-\bm{n}\cdot \rotoperator).
\end{align*}
\end{lemma}
{\startproof
As $\mathcal{J}$ commutes with the action of $\SL$, it anti-commutes with the generators. In particular it anti-commutes with
$\bm{n} \cdot \rotoperator$.  Suppose $|m\rangle$  is any eigenstate of $\bm{n} \cdot \rotoperator$
with eigenvalue $m$.  Then
\begin{align*}
(\bm{n} \cdot \rotoperator) \mathcal{J} |m\rangle
= - \mathcal{J} (\bm{n} \cdot \rotoperator) |m\rangle
= - m \mathcal{J} |m\rangle
\end{align*}
so that $\mathcal{J} |m\rangle$ is an eigenstate of $\bm{n} \cdot \rotoperator$ with eigenvalue $-m$.  Thus
\begin{align*}
\Pi_{\cal S}(\bm{n}\cdot \rotoperator)\, {\cal J}\mathcal{I} |m\rangle
=\chi_{\cal S}(-k)\; {\cal J}\mathcal{I} |m\rangle
={\cal J}\;\Pi_{\cal S}(-\bm{n}\cdot \rotoperator)\, |m\rangle.
\end{align*}
Because the states $|\lambda \rangle$ span $\hilbert_{k,p}$, the result follows.
\finishproof}
\begin{theorem}
\label{th:arbit_pos}
The proper vertex amplitude \eqref{proper_vertex} is equal to
\begin{align*}
\begin{split}
A_v^+=&(-1)^{\Xi}\int_{\SL^5}\delta(X_4)\prod_a\dif X_a\prod_{a<b}
\alpha(\mathcal{I}\psi_{ab}, X_{ab}\,\Pi_{ba}(\{X_{a'b'}\})\,\mathcal{I}\,\psi_{ba})\\
=&(-1)^{\Xi}\int_{\SL^5}\delta(X_4)\prod_a\dif X_a\prod_{a<b}
\alpha(\Pi_{ab}(\{X_{a'b'}\})\,\mathcal{I}\psi_{ab},X_{ab}\,\mathcal{I}\,\psi_{ba})\\
=&(-1)^{\Xi}\int_{\SL^5}\delta(X_4)\prod_a\dif X_a\prod_{a<b}
\alpha(\mathcal{I}\,\Pi_{ab}(\{X_{a'b'}\})\,\psi_{ab},X_{ab}\,\mathcal{I}\,\psi_{ba})
\end{split}
\end{align*}
\end{theorem}
{\startproof
This is proven by successively applying lemma \ref{lemma9}, lemma \ref{lemma10} and lemma \ref{lemma13} to \eqref{proper_vertex}, and using the relation between $\alpha(\cdot, \cdot)$ and $(\cdot , \cdot)$ and the fact that
\begin{align*}
\tr(\pauli_i X_{ab} X_{ab}^\dagger)\rotoperator^i
= \epsilon_{IJKL} \timeo^I (\widehat{X}_{ab} \timeo)^J \hat{J}^{KL} .
\end{align*}

\finishproof}

\section{Relation of the expression for EPRL used here to that of Barrett \textit{et al.}}
\label{convapp}

\subsection{Summary}

In {\asymptoticL}, the expression for the vertex amplitude is given as
\begin{align}
\label{oldA}
\tilde{A}_v(k, \xi)
:= (-1)^{\Xi}\int_{\SL^5} \delta(X_0) \prod_{a=0}^{4} \dif X_a
\prod_{a<b} \tilde{\mathcal{P}}_{ab}(\xi, X)
\end{align}
with propagators
\begin{align}
\label{oldProp}
\tilde{\mathcal{P}}_{ab}(\xi, X):= \alpha(\overline{X}_a \mathcal{I} \phi_{\xi_{ab}}^{k_{ab}},
\overline{X}_b \mathcal{I} \phi_{\xi_{ba}}^{k_{ab}})
\end{align}
where
\begin{align}
\label{phicohdef}
\phi^k_{\xi}(z):=\sqrt{\frac{d_k}{\pi}}[z,\xi]^{2k}
:= \sqrt{\frac{d_k}{\pi}}(z_0 \xi_1 - z_1 \xi_0)^{2k}
= \sqrt{\frac{d_k}{\pi}}(-1)^{2k}\langle J\xi, z\rangle^{2k}.
\end{align}
The expression (\ref{oldA}), (\ref{oldProp}) has the following two disadvantages:
(1.) The complex conjugation of $X$ in the expression leads to an unusual
relationship between the bivector conjugate to $X$ and the generators of the
the Lorentz group acting in the irreducible representations\footnote{If
one understands the expression for the vertex amplitude as the evaluation of the BF amplitude for a `simple' $\SL$ boundary state
as in (\ref{eprlFromBF}), one can derive the relationship between
the quantization of the bivectors conjugate to $X$ and the Lorentz generators on the
irreps.  Due to $X$ being complex conjugated in the expression, one finds that they are not equal, but are related by the adjoint action of reflection about $y=0$.}, and
(2.)
The operator equation satisfied by each of the coherent states $\phi_\xi^k(z)$ seems to
be in conflict with the geometrical interpretation of its label, in that
$\phi_{\xi}^{k}(z)$ is a maximum-eigenvalue eigenstate of $n \cdot L$
for $n = n_{J \overline{\xi}}$ rather than $n=n_{\xi}$ (see theorem \ref{othercoh}).

As the interpretation of these structures are central to the correct quantization
of the Einstein-Hilbert condition as well as the analysis of the resulting asymptotics, we have opted to use the following more straight-forward expression for the
vertex amplitude:
\begin{align}
\label{newA}
A_v(k, \xi)
:= (-1)^{\Xi} \int_{\SL^5} \delta(X_0)\left(\prod_{a=0}^{4} \dif X_a \right)
\prod_{a<b} \mathcal{P}_{ab}(\xi, X)
\end{align}
with different propagators
\begin{align*}
\mathcal{P}_{ab}(\xi, X):= \alpha(X_a \mathcal{I} C_{\xi_{ab}}^{k_{ab}},
X_b \mathcal{I} C_{\xi_{ba}}^{k_{ab}}).
\end{align*}
It is easy to check (see theorem \ref{otherprop}) that
\begin{align}
\label{Xrelation}
\mathcal{P}_{ab}(\xi, X) = \tilde{\mathcal{P}}_{ab}(\xi, (X^\dagger)^{-1})
\end{align}
so that the two expressions (\ref{oldA}) and (\ref{newA}) for the vertex amplitude
are simply related by a change of the integration
variables $X$ which preserves the Haar measure, so that in fact
\begin{align*}
\tilde{A}(k,\xi) = A(k,\xi).
\end{align*}
Thus, the vertex amplitude here is not different from that in {\asymptoticL}, but only the expression differs.
Because of this,
it is immediate that the asymptotics of the vertex amplitude are the same for the two different conventions.
Nevertheless, intermediate equations will change.  To assist in translating the intermediate
equations, let $\tilde{X}$ denote the group variables in the framework of {\asymptoticL},
distinguishing them from
the group variables $X$ used in the present paper. By requiring equality of the vertex amplitude
integrands,
\begin{align*}
\prod_{a<b} \mathcal{P}_{ab}(\xi, X) = \prod_{a<b} \tilde{\mathcal{P}}_{ab}(\xi, \tilde{X}),
\end{align*}
one is lead to
\begin{align}
\label{mapping}
\tilde{X} \equiv (X^\dagger)^{-1}.
\end{align}
The physical bivectors as defined in our conventions (\ref{eqn:geomB}, \ref{eqn:contB})
(to ensure we are talking about the same quantity),
but reconstructed from $\tilde{X}$ as in {\asymptoticL} are given by
\begin{align}
\label{Btdef}
\tilde{B}_{ab} := \widehat{\tilde{X}}_a \triangleright b_{ab}
\equiv -\gamma \kappa k_{ab} \widehat{\tilde{X}}_a \triangleright \timeo \wedge (0, \bm{n}_{\xi_{ab}})
\end{align}
that is, the same expression (\ref{eqn:Brecon})
that $B_{ab}$ takes in terms of $X_a, \xi_{ab}$
in the present paper.  Equations (\ref{mapping}),(\ref{Btdef}), and (\ref{eqn:Brecon}) then imply
(see theorem \ref{otherrecon})
\begin{align}
\label{Brelation}
\tilde{B}_{ab} = - P \triangleright B_{ab}
\end{align}
where $P$ denotes the spatial parity operator negating all spatial components.
To set up the asymptotic problem in section \ref{sssec:eprlcoh}, one must expand the propagator
$\mathcal{P}_{ab}$ as an integral over
$\mathbb{CP}^1$.  It is easiest to do this by using equation (\ref{Xrelation}) and the expression for $\tilde{P}_{ab}$
given in section 3.3 of {\asymptoticL}, which yields
%
%
\begin{align*}
{\cal P}_{ab} &= c_{ab} d_{k_{ab}} \int_{\mathbb{CP}^1}
\left\langle X_a^{-1} z_{ab}, X_a^{-1}z_{ab}\right\rangle^{-1-ip_{ab}-k_{ab}}
\left\langle X_a^{-1} z_{ab}, \xi_{ab} \right\rangle^{2k_{ab}}\\
& \hspace{2.5cm} \left\langle X_b^{-1} z_{ab}, X_b^{-1}z_{ab}\right\rangle^{-1+ip_{ab}-k_{ab}}
\left\langle J \xi_{ba}, X_b^{-1} z_{ab} \right\rangle^{2k_{ab}} \Omega_{z_{ab}}
\end{align*}
This expression then defines for us $z_{ab}$. If we express the vertex amplitude
(\ref{eqn:vertexampl}) as an integral (\ref{eqn:vertexampl2})
over the $X$'s \textit{and} the $z$'s, then the integrand in (\ref{eqn:vertexampl2}) and the integrand in {\asymptoticL}
will again be related simply by the replacement $\tilde{X} \mapsto X = (\tilde{X}^\dagger)^{-1}$, without changing
$z$ or $\xi$. From this equality of the vertex integrands via the translation
(\ref{mapping}), it follows that the critical points in the two frameworks will again be equal via (\ref{mapping}).
The consequences of the critical point equations for $\tilde{B}_{ab}$ are simply
the weak bivector geometry constraints;  as these constraints
are invariant under the transformation (\ref{Brelation}),
the critical point equations for $B_{ab}$ again will be simply
the weak bivector geometry constraints.  Furthermore, the relation
(\ref{Brelation}) implies that, at the critical points, both
$B_{ab}$ and $\tilde{B}_{ab}$ are in the same $\mu$ sector and determine the same
Regge geometry, so that the identification of the different terms in the asymptotics with different values of $\mu$ is the same
in the two frameworks.

In summary, the two frameworks are fully equivalent, and to translate the equations in {\asymptoticL}
to the conventions used here (besides the change in the definition of the bivectors
(\ref{eqn:geomB}, \ref{eqn:contB})), only one rule is necessary:
\begin{quote}
\textit{\textbf{$X$ is everywhere replaced by $(X^\dagger)^{-1}$ except in the expression for reconstructing $B_{ab}$.}}
\end{quote}
In the following section we prove some of the statements claimed above.

\subsection{Proofs}

\begin{theorem}
\label{othercoh}
The coherent states $\phi^k_{\xi}(z)$ satisfy
$\left(\bm{n}_{J\overline{\xi}}\cdot\rotoperator\right) \phi^k_{\xi}= k\phi^k_{\xi}$ .
\end{theorem}
{\startproof
From equations (\ref{Ccohdef}) and (\ref{phicohdef}), one has
\begin{align}
\label{xif}
\phi^k_\xi(z) = (-)^{2k} C^k_{J\bar{\xi}}(z).
\end{align}
The result then follows from the properties of $C^k_{J\bar{\xi}}$.
\finishproof}

As an aside which will be used below, note that
\begin{align}
\label{yrot}
J\bar{\xi} = r_y(-\pi) \xi,
\end{align}
where
\begin{displaymath}
r_y(-\pi):=e^{-\frac{i\pi}{2}\pauli_2} = -i \pauli_2 =
\left( \begin{array}{cc} 0 & -1 \\ 1 & 0 \end{array}\right)
\end{displaymath}
is the $\SU(2)$ element corresponding to a rotation about the $y$-axis by $-\pi$.

\begin{theorem}
\label{otherprop}
\begin{align*}
\mathcal{P}_{ab}(\xi, X) = \tilde{\mathcal{P}}_{ab}(\xi, (X^\dagger)^{-1})
\end{align*}
\end{theorem}
{\startproof
\begin{align*}
\tilde{\mathcal{P}}_{ab}(\xi, (X^\dagger)^{-1}) &=  \alpha((X_a^T)^{-1} \mathcal{I} \phi_{\xi_{ab}}^{k_{ab}},
(X_b^T)^{-1} \mathcal{I} \phi_{\xi_{ba}}^{k_{ab}})
=  \alpha((X_a^T)^{-1} \mathcal{I} C_{r_y(-\pi)\xi_{ab}}^{k_{ab}},
(X_b^T)^{-1} \mathcal{I} C_{r_y(-\pi)\xi_{ba}}^{k_{ab}}) \\
&=  \alpha((X_a^T)^{-1} r_y(-\pi) \mathcal{I} C_{\xi_{ab}}^{k_{ab}},
(X_b^T)^{-1} r_y(-\pi) \mathcal{I} C_{\xi_{ba}}^{k_{ab}}) \\
&=  \alpha( r_y(-\pi) X_a \mathcal{I} C_{\xi_{ab}}^{k_{ab}},
 r_y(-\pi) X_b  \mathcal{I} C_{\xi_{ba}}^{k_{ab}}) \\
 &=  \alpha(X_a \mathcal{I} C_{\xi_{ab}}^{k_{ab}}, X_b  \mathcal{I} C_{\xi_{ba}}^{k_{ab}})
 = \mathcal{P}_{ab}(\xi, X)
\end{align*}
where (\ref{xif}) and (\ref{yrot}) were used in the first line,
the $SU(2)$ covariance of $\xi \mapsto C_\xi^k$ and $\mathcal{I}$ were used in the second line,
the expression for the inverse of an $\SL$ matrix, $X^{-1} = r_y(-\pi) X^T r_y(\pi)$ was used in the third line,
and the $\SL$-invariance of $\alpha$  in the last line.
\finishproof}

\begin{lemma}
\label{barXlemm}
For any $X \in \SL$,
\begin{align*}
\widehat{\overline{X}} = P_y \widehat{X} P_y
\end{align*}
where $P_y$ is the parity operator flipping the sign of the $y$ component.
\end{lemma}
{\startproof
For any $Y \in \SL$ and $V^I \in \R^4$, by definition
\begin{align*}
\rho(\widehat{Y}V) &= Y \rho(V) Y^\dagger \\
(\widehat{Y}V)^0 + \pauli_i (\widehat{Y}V)^i &=
Y(V^0 + \pauli_i V^i) Y^\dagger,
\end{align*}
so that
\begin{align*}
(\widehat{Y}V)^0 = \frac{1}{2} \tr\left[Y(V^0 + \pauli_i V^i)Y^\dagger\right] \qquad \text{and} \qquad
(\widehat{Y}V)^i = \frac{1}{2} \tr \left[\pauli_i Y(V^0 + \pauli_j V^j)Y^\dagger \right].
\end{align*}
Thus
\begin{align*}
(\widehat{\overline{X}}V)^0 &= \frac{1}{2} \tr\left[\overline{X}(V^0 + \pauli_i V^i)X^T\right]
= \frac{1}{2} \tr\left[X(V^0 + (\pauli_i)^T V^i)X^\dagger\right]
= \frac{1}{2} \tr\left[X(V^0 + (P_y)^j{}_i\pauli_j V^i)X^\dagger\right] \\
&= \frac{1}{2} \tr\left[X(V^0 + \pauli_j (P_y V)^j)X^\dagger\right]
= (\widehat{X}P_y V)^0 =  (P_y \widehat{X}P_y V)^0
\end{align*}
where the invariance of the trace under transpose was used in the second equality, and the explicit expression
for the Pauli matrices in the third equality.
Furthermore,
\begin{align*}
(\widehat{\overline{X}}V)^i &= \frac{1}{2} \tr\left[\pauli_i \overline{X}(V^0 + \pauli_j V^j)X^T\right]
= \frac{1}{2} \tr\left[X(V^0 + (\pauli_j)^T V^j)X^\dagger (\pauli_i)^T \right] \\
&= (P_y)^i{}_k \frac{1}{2} \tr\left[\pauli^k X(V^0 + \pauli_j (P_y V)^j)X^\dagger \right]
= (P_y)^i{}_k (\widehat{X}P_y V)^k = (P_y \widehat{X} P_y V)^i .
\end{align*}
\finishproof}

\begin{theorem}
\label{otherrecon}
\begin{align*}
\tilde{B}_{ab} = - P \triangleright B_{ab}
\end{align*}
\end{theorem}
{\startproof
\begin{align*}
\tilde{B}_{ab} &:= \widehat{\tilde{X}}_a \triangleright b_{ab}
 := \widehat{(X_a^\dagger)^{-1}} \triangleright b_{ab}
 = \widehat{r_y(\pi)\overline{X}_ar_y(-\pi)} \triangleright b_{ab}
 = \left(R_y(\pi)\widehat{\overline{X}}_a R_y(-\pi)\right) \triangleright b_{ab} \\
 &= \left(R_y(\pi)P_y \widehat{X}_a P_y R_y(-\pi)\right) \triangleright b_{ab}
 = \left(P \widehat{X}_a P \right) \triangleright b_{ab}
 = - \left(P \widehat{X}_a \right) \triangleright b_{ab} = - P B_{ab}
\end{align*}
where $R_y(\theta)$ denotes rotation about the $y$-axis by angle $\theta$.
Here $X^{-1} = r_y(\pi) X^T r_y(-\pi)$ was used in the third equality, lemma \ref{barXlemm} in the fifth equality,
and the fact that $P \triangleright b_{ab} = -b_{ab}$ in the penultimate equality.
\finishproof}

%
%

%

\end{document}